\documentclass[fleqn,useAMS]{mnras}
\usepackage{mathptmx}
\usepackage[T1]{fontenc}
\usepackage{ae,aecompl}
\usepackage{graphicx}	% Including figure files
\usepackage{amsmath}	% Advanced maths commands
\usepackage{amssymb}	% Extra maths symbols
\usepackage{xspace}
\usepackage{color}
\usepackage{txfonts}

\newcommand {\bc}{\begin {center}}
\newcommand {\ec}{\end {center}}
\newcommand {\be}{\begin {equation}}
\newcommand {\ee}{\end {equation}}

\newcommand{\gx}{GX~301\mbox{--}2\xspace}
\def\hxmt {\textit{Insight}-HXMT}
\def\xte {\emph{RXTE}}
\newcommand{\exosat}{\textsl{EXOSAT}\xspace}
\newcommand{\batse}{\textsl{BATSE}\xspace}
\def\xte {\emph{RXTE}}

\def\asca {\emph{ASCA}}
\newcommand{\integral}{\textsl{INTEGRAL}\xspace}

\newcommand{\xmm}{\textsl{XMM-Newton}\xspace}
\newcommand{\bepposax}{\textsl{BeppoSAX}\xspace}

\newcommand{\maxi}{\textsl{MAXI}\xspace}

\newcommand{\kev}{\,$\mathrm{keV}$ }

\newcommand{\pcm}{\,$\mathrm{cm^{-2}}$ }

\newcommand{\ie}{\textit{i}.\textit{e}. }

\title[\textit{Insight}-HXMT observation of GX 301--2]{Timing and spectral variability of high mass X-ray pulsar GX 301--2 over orbital phases observed by \textit{Insight}-HXMT}
\author[Y. Z. Ding and W. Wang et al.]
{\bf \small Y. Z. Ding,$^{1,2,3}$ W. Wang,$^{1,2}$\thanks{wangwei2017@whu.edu.cn} P. R. Epili,$^{1,2}$ Q. Liu,$^{1,2}$ M. Y. Ge,$^{4}$ F. J. Lu,$^{4}$ J. L. Qu,$^{4}$ L. M. Song,$^{4}$ S. Zhang,$^{4}$ S. N. Zhang$^{4,5}$ 
	\\
	$^1$ School of Physics and Technology, Wuhan University, Wuhan 430072, China  \\
	$^2$ WHU-NAOC Joint Center for Astronomy, Wuhan University, Wuhan 430072, China \\
	$^3$ Hongyi Honor College, Wuhan University, Wuhan 430072, China \\
	$^4$ Key Laboratory of Particle Astrophysics, Institute of High Energy Physics, Chinese Academy of Sciences, Beijing 100049, China \\
	$^5$ University of Chinese Academy of Sciences, Chinese Academy of Sciences, Beijing 100049, China \\
	}

\pubyear{2021}

\begin{document}
\label{firstpage}
\pagerange{\pageref{firstpage}--\pageref{lastpage}}
\maketitle

\begin{abstract}
We report the orbital X-ray variability of high mass X-ray binary (HMXB) \gx. \gx undergone a spin up process in 2018--2020 with the period evolving from $\sim$ 685 s to 670 s. The energy resolved pulse-profiles of the pulsar in 1--60 keV varied from single peaked and sinusoidal shapes to multi-peaked across different orbital phases. Pulse fractions evolving over orbit had negative correlations with the X-ray flux. The broad-band X-ray energy spectrum of the pulsar can be described with a partial covering negative positive cutoff power-law continuum model. Near the periastron passage of the pulsar we found a strong variation in the additional column density ($NH_{2}$), which correlated with variation of the flux. Curves of growth for both Fe K$\alpha$ and Fe K$\beta$ lines were plotted to investigate the distribution of matter around neutron star. We have also found the evidence of two cyclotron absorption lines in the phase-averaged spectra in \gx, with one line of 30--42 keV and the other line varying in 48--56 keV. Both two line's centroid energies show the similar relationship with X-ray luminosity: positive correlation in lower luminosity range, and negative relation above a critical luminosity of $10^{37} \rm erg s^{-1}$. We estimated the surface magnetic field of the neutron star in \gx at $\sim (0.5-2)\times 10^{13}$ G. Two cyclotron line energies have a nearly fixed ratio of $\sim 1.63$ while having a low strength ratio ($\sim 0.05$), suggesting that these two features may actually be one line.

\end{abstract}


%%%%%%%%%%%%%%%%% BODY OF PAPER %%%%%%%%%%%%%%%%%%%%%%%%%%%%%%%%%%%%%%%%%%%%%%	

\section{Introduction}

\gx is a high mass X-ray binary system consisting of a neutron star
in a 41.5~d eccentric orbit with a donor star of early type known as Wray 977 (White \& Swank 1984). The donor star is massive ($\simeq39-53~M_{\odot}$) having a size of
$\simeq62~R_{\odot}$ and is at a distance of $\simeq 3$ kpc (Kaper, van der Meer \& Najarro 2006). Most recent \textit{Gaia} measurement implies d=$3.53^{+0.40}_{-0.52}$kpc (Bailer-Jones et al. 2018). The stellar wind of this B1 hypergiant companion star is rather dense ($\dot{M}_{w}\sim10^{-5}~M_{\odot}yr^{-1}$)  and slow ($\simeq 300~\rm km~s^{-1}$).
This high stellar wind mass loss rate could propel the accretion rate onto
the neutron star to an observed luminosity of $L_{x}\sim10^{37} \rm ~erg~s^{-1}$.
The $5\times10^5~L_{\odot}$ luminosity of the Wolf-Rayet companion
Wray 977 (Kaper et al. (2006) makes it among the most luminous stars in the Galaxy.
The neutron star (NS) in this system has the mass of $1.85\pm 0.60 M_{\odot}$ based on radial-velocity studies (Kaper et al. 2006) and is a rather slowly rotating pulsar
($P_{spin}\simeq 680~\rm s$, Nabizadeh et al. 2019; Abarr et al. 2020; M{\"o}nkk{\"o}nen et al. 2020), compared to majority of accreting X-ray pulsars.

The neutron star in \gx shows strong orbital variability as it traverses through
its binary orbit. One of the hallmark characteristics of this system is the
recurrence of a pre-periastron flare at the NS. A densely emanating gas stream
from the companion hypergiant Wray 977 is recurrently intercepted by the NS
shortly before every periastron passage (Leahy 1991; Leahy 2002; Leahy \& Kostka 2008). It results in intense X-ray flares with a 25\% increase in the X-ray luminosity (Pravdo et al. 1995; Pravdo \& Ghosh2001).  It is also evident that the continuum X-ray emissions of \gx modulated by the orbital period attain a peak around the orbital phase 1.4~d before the periastron (see Fig. 3 of Sato et al. 1986).

%%%%%%%%%% Orbital Variability %%%%%%%%
Leahy \& Koska (2008) have studied a ten year long orbital light curve of \gx
with \xte. These studies reveal that a simple model consisting of a gas stream
plus stellar wind accretion onto the NS could explain the substantial changes
in the orbital light curve suggesting bright, medium and dim intensity levels across the orbit. The orbital phase dependence of the X-ray flare and flux variation
in \gx has been studied previously with \exosat (Haberl 1991), CGRO/\batse
(Koh et al. 1997), \asca \ (Endo et al. 2002), \xte/ASM (Leahy 2002), \xte/PCA (Mukherjee \& Paul  2004), \bepposax (La Barbera et al. 2005), and recently with \xmm and \maxi (Islam \& Paul 2014), \integral (Doroshenko et al. 2010; Yu \& Wang 2016).

The broad-band X-ray spectrum of \gx can be explained by various phenomenological models with a power law modified by cutoff at higher energy. At lower energy the spectrum of \gx gets heavily absorbed (F\"urst et al. 2011). This is due to the presence of variable column density, $N_{H}$ (in the range of $10^{23} - 2\times10^{24}$~\pcm ) along
the observers line of sight throughout the orbit, and thus indicating the
accretion from clumpy stellar wind. To explain the low-energy part of the spectrum, previous studies (La Barbera et al. 2005; Mukherjee \& Paul 2004) also used a partial covering component apart from the usual galactic absorption. At energies 6-7\kev, \gx also shows many fluorescent lines (F\"urst et al. 2011; Zheng et al. 2020) with Fe~K$\alpha$ emission line at 6.4\kev being very bright and prominent at different orbital phases.  From the \asca  observations spanning across intermediate, apastron \& periastron orbital phases of \gx, Endo  et al. (2002) noted a larger width of iron K$\alpha$ line (i.e 40--80 eV). A pulsed Fe K$\alpha$ line emission was also reported by Chandra observation near periastron (Liu et al. 2018).

In the accreting magnetized neutron star systems, the accretion flow is directed by magnetic field towards the small regions located close to the magnetic poles, where the kinetic energy of the material turns into heat, being emitted in the form of X-ray radiation. A strong magnetic field in the vicinity of NS surface quantizes the energy of electrons and consequently, the Compton scattering and free-free absorption become resonant at the cyclotron energy
\be
E_{\rm cyc}\approx 11.6\,\left(\frac{B}{10^{12}\,{\rm G}}\right)\,{\rm keV},
\ee
and its harmonics, where $B$ is the field strength. The resonances result in cyclotron scattering features in the energy spectra of X-ray pulsars. At present, there are more than 30 X-ray pulsars with detected cyclotron scattering features (Staubert et al. 2019), with the fundamental cyclotron line energies from $\sim 9-90$ keV (Wang 2014). In most of them, only one cyclotron absorption line is reported.

The cyclotron resonant scattering feature (CRSF) of \gx around 40 keV was first discovered by Makishima \& Mihara (1992) based on Ginga observations. However, the cyclotron line centroid energy in \gx shows large variations and deviations by different measurements. Based on RXTE data, Kreykenbohm et al. (2004) reported the variable cyclotron line with energies from $30- 40$ keV according to the phase-resolved spectra. La Barbera et al. (2005) detected the CRSFs with the centroid energies from $45- 55$ keV in different orbital phases with BeppoSAX observations. Doroshenko et al. (2010) detected a cyclotron scattering line at $\sim 50$ keV in \gx with INTEGRAL observations. Suzaku broadband spectroscopy on \gx showed the variations of CRSFs from $\sim 30- 45$ keV (Suchy et al. 2012). Yu \& Wang (2016) reported a variation of the line energy from $\sim 36- 48$ keV with INTEGRAL/IBIS data. With high resolution spectroscopy by Nustar, F\"urst et al. (2018) claimed that this reported broad feature is in fact two absorption lines, at $\sim 35-40$ keV and $\sim 50-55$ keV separately.

The Chinese hard X-ray modulation telescope (\textit{Insight}-HXMT, Zhang et al. 2020) provides the high timing resolution and broad X-ray spectral studies on X-ray binaries. Since its launch on 15 June 2017, \textit{Insight}-HXMT carried out frequent pointing observations on the source GX~301-2, covering the different orbital phases, which provided a chance for us to detailedly study the variations of pulse emission and spectral properties versus orbital phase. In \S 2, the \textit{Insight}-HXMT observations and data analysis procedure are introduced. The pulse profiles and the characteristics are studied in \S 3, and variations of the spectral properties with different orbital phases are shown in \S 4. We discussed spectral variations and CRSFs in \S 5. Summary and conclusion are presented in \S 6.

\section{X-ray Observations and data analysis}
%% From ApJ
\textit{Insight}-HXMT consists of three main detectors: the Low Energy X-ray Detector (LE, Chen et al. 2020), the Medium Energy X-ray Telescope (ME, Cao et al. 2020), and  the High Energy X-ray Telescope (HE, Liu et al. 2020). HE contains 18 cylindrical NaI(Tl) / CsI(Na) detectors with a total detection
area of 5000 cm$^2$ in the energy range of 20--250 keV. ME is composed of 1728 Si-PIN detectors with a total detection area of 952 cm$^2$ covering the range of 5--30 keV. LE uses Swept Charge Device (SCD) with a total detection area of 384 cm$^2$ covering 1--15 keV. \textit{Insight}-HXMT have carried out a series of performance verification tests by observing blank sky, standard sources, showing good calibration state and estimation of the instrumental background (Li et al. 2020).

\begin{figure}
	\includegraphics*[width=\columnwidth]{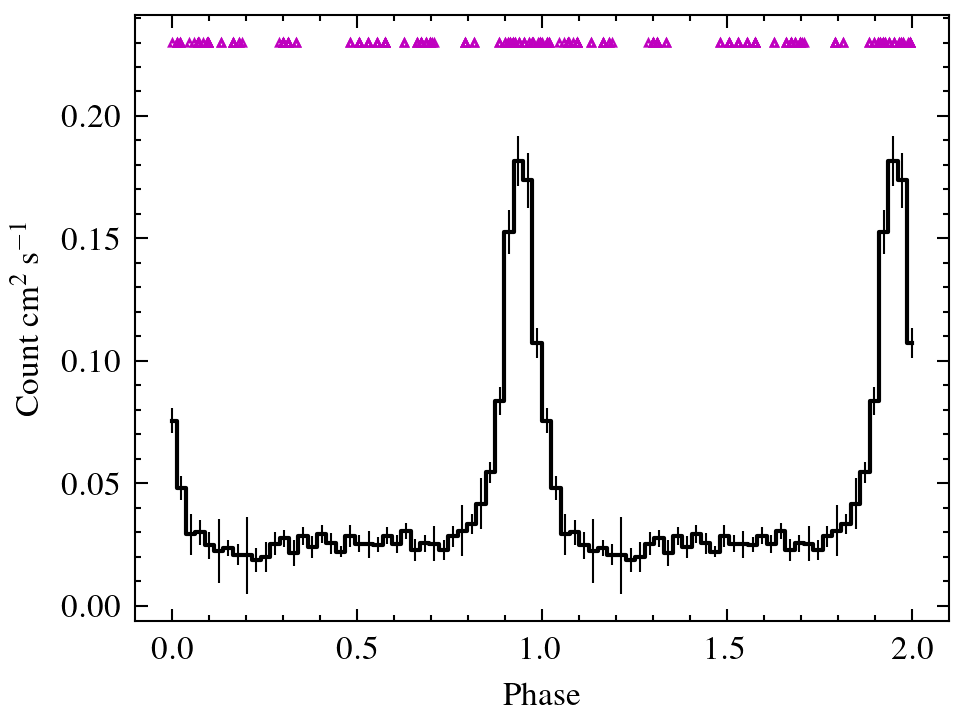}
	\caption{\textit{Insight}-HXMT observation log (magenta triangle on top) over-plotted on the Swift/BAT orbital phase folded count rate. Orbital period and periastron passage are from Doroshenko et al. (2010). HXMT data covered most of the orbital phases. We label the phases 0.2--0.9 as the {\em apastron}, and other phases as {\em periastron}. }
	\label{BATall}
\end{figure}

\gx has been observed numerously by HE, ME and LE telescopes onboard \textit{Insight}-HXMT satellite between August~3$^{rd}$,~2017 and June~3$^{rd}$,~2020 (i.e MJD: 57968 -- 59003). These observations cover different orbital phases of the binary orbit (see Figure \ref{BATall}). A log of observations (information of OBSIDs) used in this paper is given in the table as the supplementary material. It should be noted that the LE telescope data of observations with OBSID:P0101309 have been analyzed and published by Ji et al. (2021), who have studied the source at 5.5--8.5 keV. In the following science analysis, we filtered data with the criterion: (1) pointing offset angle $< 0.1^\circ$; (2) pointing direction above Earth $> 10^\circ$; (3) geomagnetic cut-off rigidity value $> 8$ GeV; (4) time since SAA passage $> 300$ s and time to next SAA passage $> 300$ s; (5) for LE observations, pointing direction above bright Earth $> 30^\circ$. The \textit{Insight}-HXMT Data Analysis Software Package (HXMTDAS) V2.02 was used in this work.

\begin{figure}
\includegraphics[width=\columnwidth]{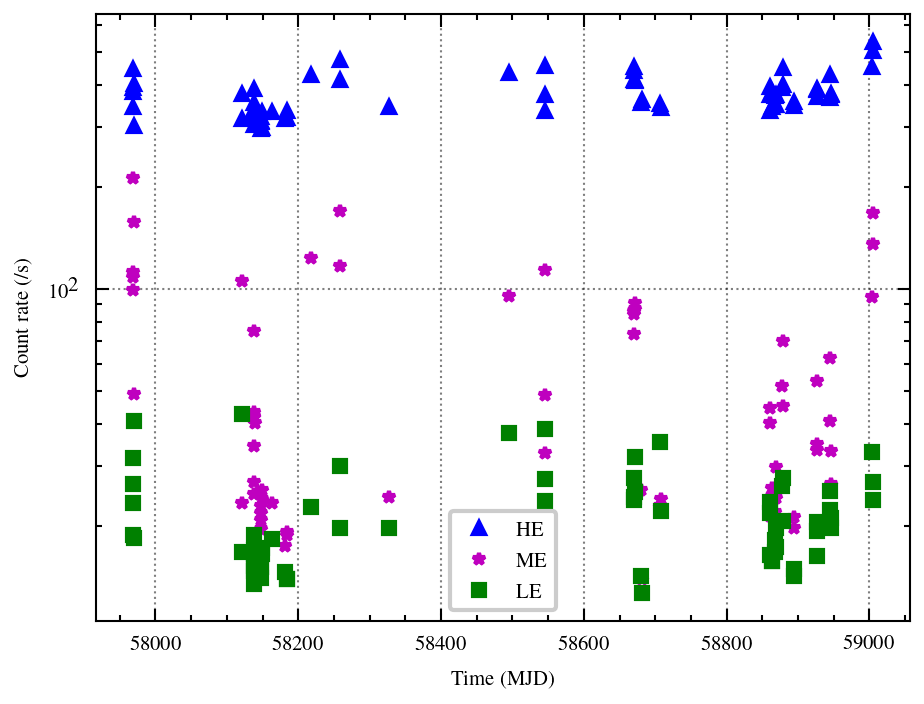}
\caption{Count rates of \gx observed by three detectors of \textit{Insight}-HXMT from August 2017 to June 2020. }
\label{fig:countrate}
\end{figure}

We obtained the X-ray light curves of \gx in three energy bands: 1--10 keV, 10--30 keV and 30--60 keV (see Figure \ref{fig:countrate}). The spin period of the neutron star in \gx is about 683 s. We made the barycentric correction of the light curves using the tool {\em hxbary}. For timing analysis, we corrected the photon arrival time for the binary motion, using ephemeris by  Doroshenko et al. (2010). In the spectral analysis, we used well-calibrated energy bands of LE, ME and HE: 3--8.5 keV, 10--30 keV and 30--70 keV, respectively (Li et al. 2020). However, ME data in 21--24 keV was ignored due to the calibration uncertainties in this energy range. All uncertainties reported in this paper are at 90\% confidence level. In the analysis and science presentation, we arranged all observations into two groups based on the orbital phase (Figure \ref{BATall}). For those having orbital phases in 0.2--0.9, we labeled them as {\em apastron} observations while others were labeled as {\em periastron}.

\section{Timing Analysis and Pulse Profiles}

\begin{figure}
\includegraphics[width=\columnwidth]{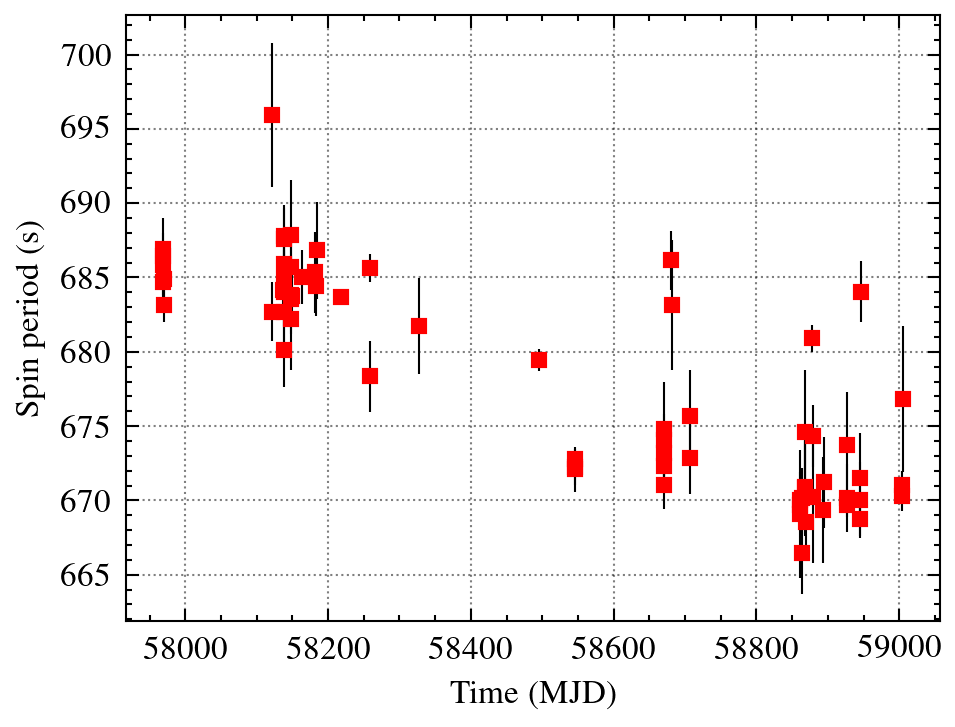}
\caption{The orbital demodulated spin period of the neutron star in \gx determined by \textit{Insight}-HXMT. Orbital parameters are from Koh et al. (1997) and Doroshenko et al. (2010). The neutron star showed a long term spin evolution. From July 2018 to June 2020, the spin period evolved from 685 s to 670 s.}
\label{fig:spinperiod}
\end{figure}

The temporal analysis aims to derive the spin period of the neutron star in \gx at different observations. Then, one can obtain the pulse profiles for three energy bands (1--10~\kev, 10--30~\kev $\&$ 30--60~\kev). The periodic signal was searched by folding the light curves, determining the chi-square ($\chi^2$) of the folded light curve and the $\chi^2$ values versus the periods. The {\em efsearch} (a built-in function in HEAsoft) was used to search for the maximum of $\chi^2$ values. We have fitted the distribution with the theoretical distribution function given by Leahy et al. (1987). We have noticed that the sampling rate of our data was uneven. Consequently, the error was estimated from Equation 2 of Larsson (1996):
\begin{equation}
\sigma_f=\frac{\sqrt{2}a\sigma_{tot}}{\sqrt{N}AT}
\end{equation}
where the parameter \textit{a} was taken to be the value derived by Monte Carlo simulation (i.e. 0.469). N is the total number of data points, A is the sinusoidal amplitude and T is the total time length for the data. The intrinsic spin period values of \gx in several OBSIDs are presented in Figure \ref{fig:spinperiod}. The spin period of \gx varied with time, reflecting the typical behavior in the wind accretion system. In addition, from 2017 -- 2020, the neutron star of \gx showed a long-term spin up from the period of $\sim 685$ s to 670 s.

\begin{figure*}
	\includegraphics[scale=0.45,angle=-90]{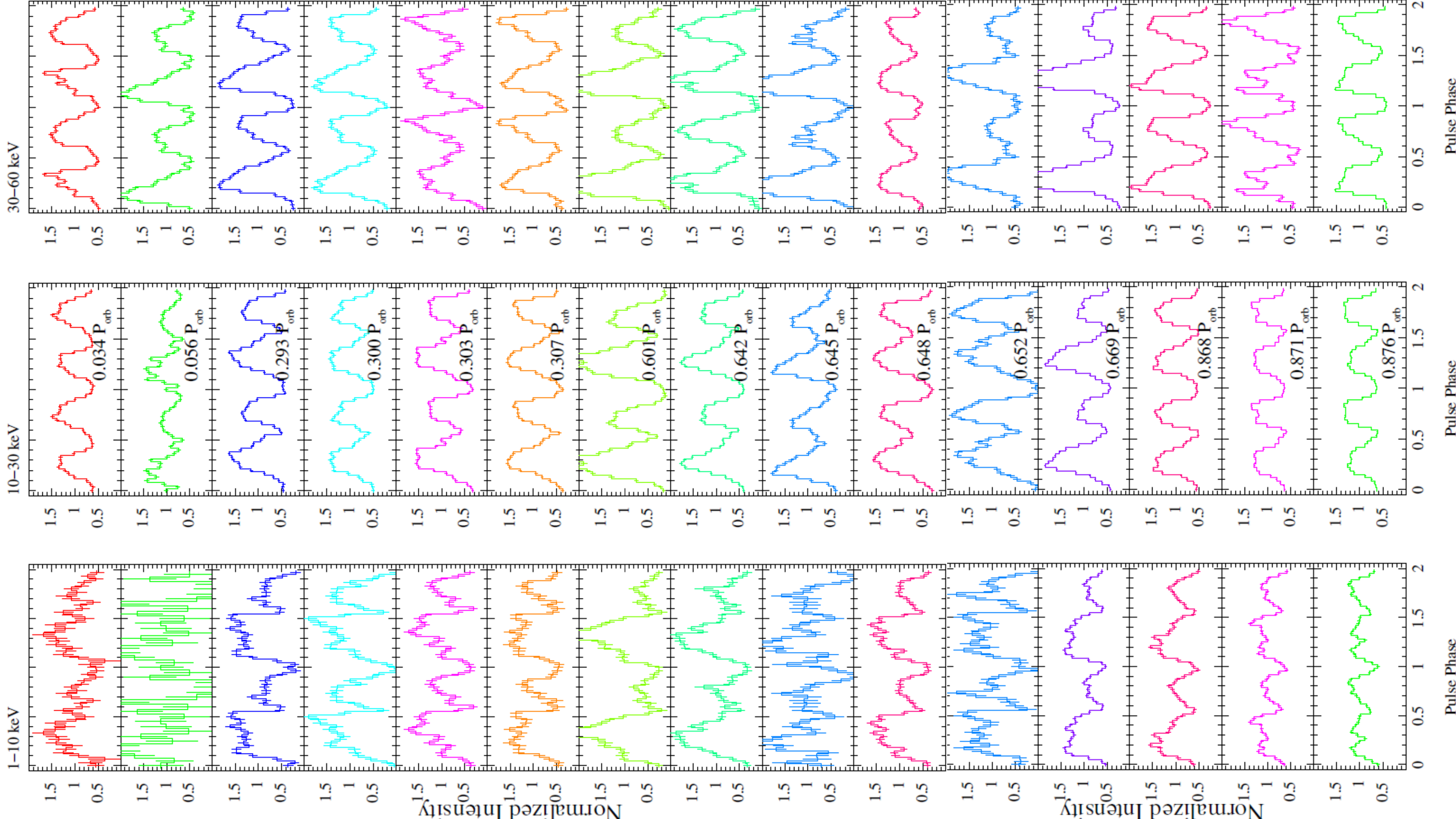}
	\includegraphics[scale=0.42,angle=-90]{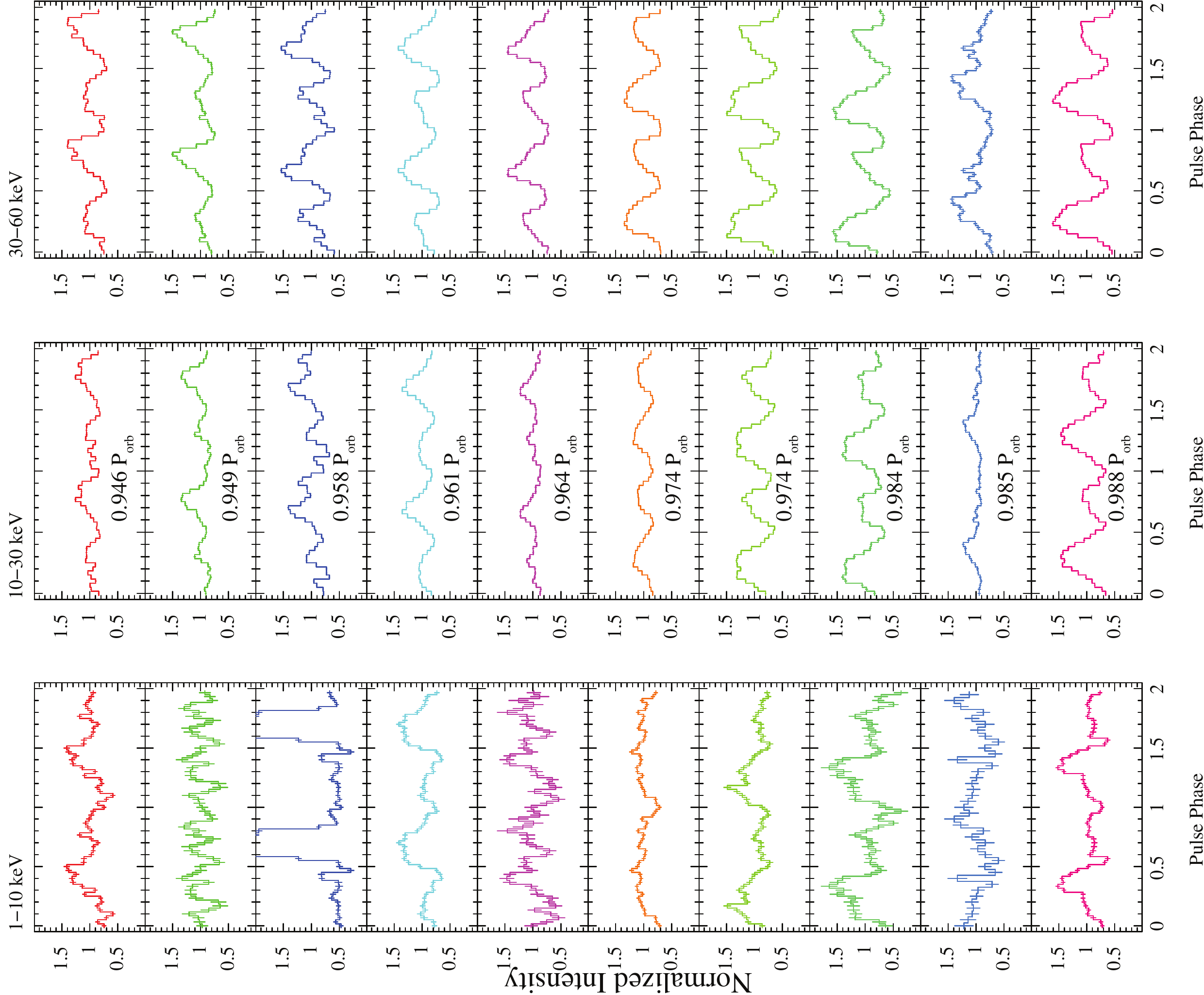}
\caption{Background subtracted pulse profiles of \gx versus orbital phases. We showed pulse profiles in three energy bands: 1--10~\kev, 10--30~\kev $\&$ 30--60~\kev, and the evolution of the profiles with the orbital phases from 0.03 to 0.99. See the text for details. }
	\label{fig:profile}
\end{figure*}

The pulse profiles of \gx are obtained in three energy bands (\ie 1--10~\kev, 10--30~\kev $\&$ 30--60~\kev) from observations of LE, ME \&
HE detectors respectively. There are 25 good OBSIDs in which at least one complete pulse profile could be constructed without any gaps. These pulse profiles are shown in
Figure~\ref{fig:profile}.

\begin{figure*}
	\includegraphics*[scale=0.6,angle=0]{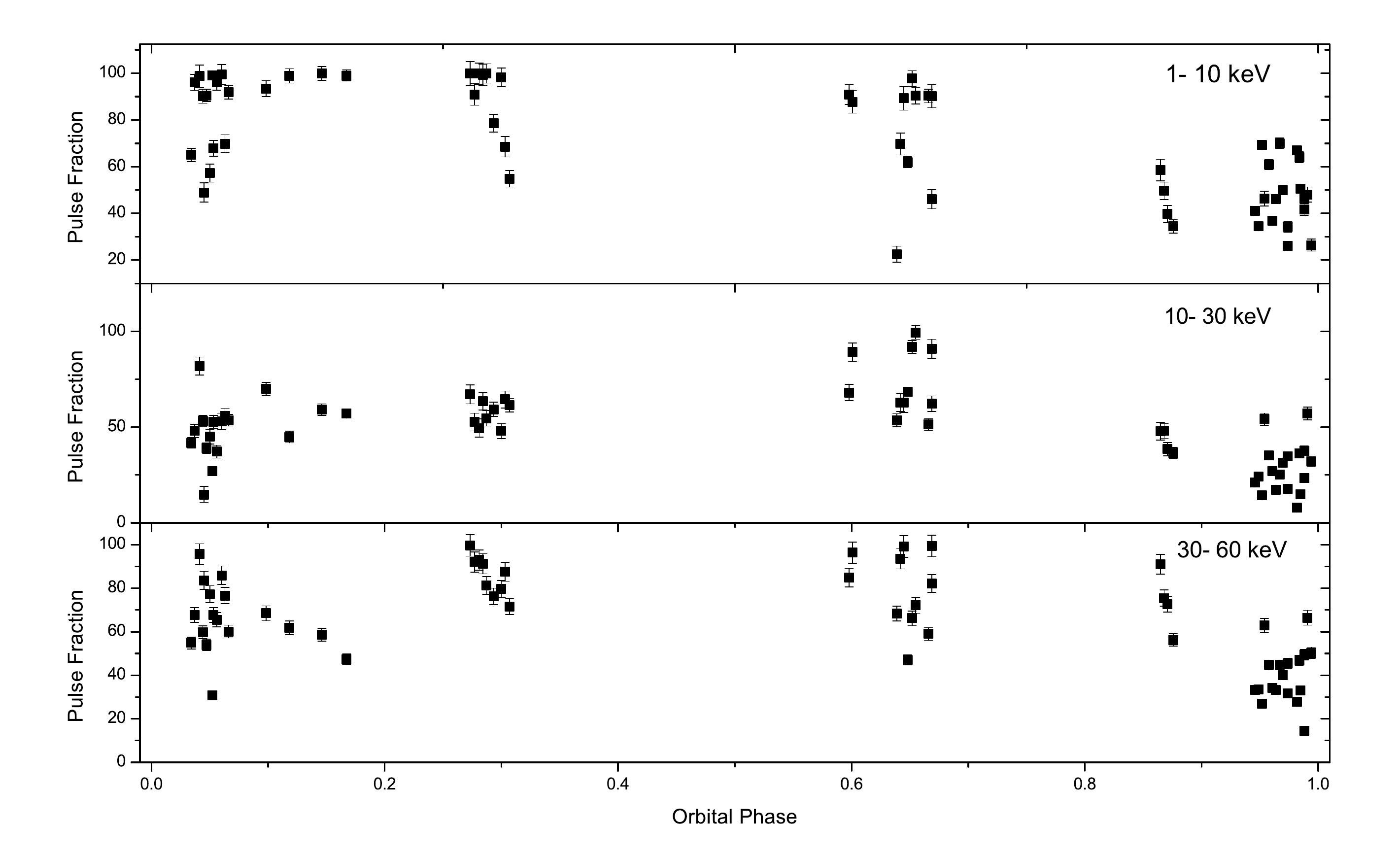}
\caption{Pulse fractions in three energy bands versus the orbital phase.}
\label{fig:pf-orbit}
\end{figure*}

\begin{figure*}
	\includegraphics*[scale=0.6,angle=0]{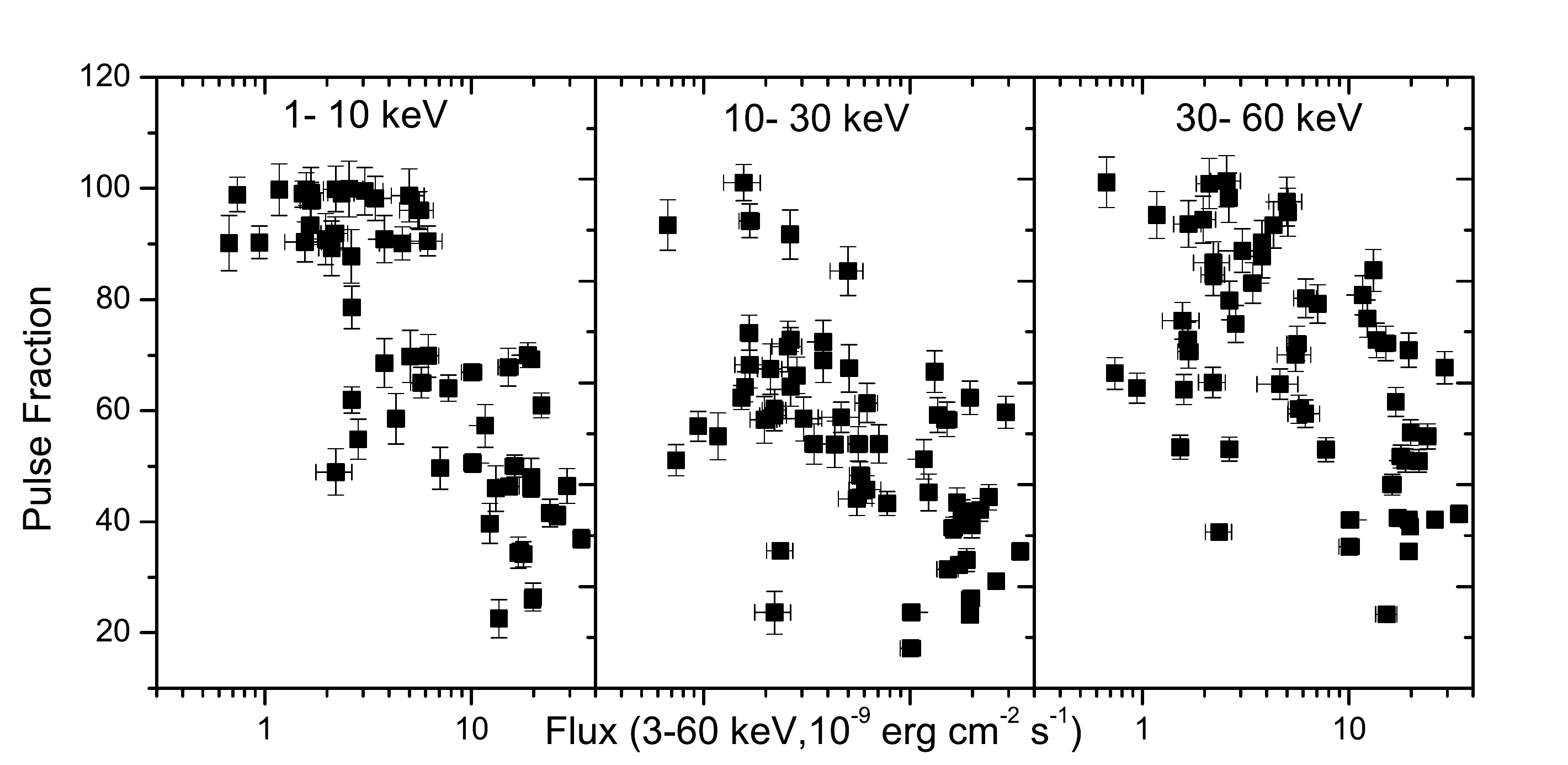}
\caption{The relationships between pulse fraction and X-ray flux. In three energy bands, there exists the negative correlation between pulse fraction and X-ray flux.}
\label{fig:pf-flux}
\end{figure*}

The pulse profiles changes with the energy bands. In hard X-ray bands above 10 keV, the neutron star generally shows double peaks, sometimes as sinusoidal shapes. In low energy band, the mini-peaks appeared in all orbital phases, which made the double peak features disappear in some orbital phases. The energy resolved pulse-profiles of pulsar are also seen to be varying from single peaked and sinusoidal shapes to multi-peaked across different orbital phases. To study the pulse profile properties over the orbital phase, we calculated the pulse fraction (defined as $PF={I_{max}-I_{min}\over I_{max}+I_{min}}$) for all light curves and show the variation of $PF$ versus orbital phase in Figure~\ref{fig:pf-orbit}. It is interesting that the mean pulse fraction at the orbital phase 0.85 --0.99 was generally lower than that in other orbital phase for three energy bands, which is also consistent with the results reported by Endo et al. (2002) and Nabizadeh et al. (2019). The neutron star system approaching the periastron showed the X-ray flares, so that the pulse fraction may depend on the X-ray luminosity. In Figure~\ref{fig:pf-flux}, we also plotted the diagrams of three pulse fractions versus observed X-ray flux from 3 --70 keV. The pulse fractions in three energy bands show negative correlation with X-ray flux, higher X-ray flux with the lower pulse fraction values.

\section{Spectroscopy over orbital phases}

\textit{Insight}-HXMT pointing observations covered different orbital phases of \gx (see Figure~\ref{BATall}). To understand the spectral properties and variations over the orbit, we obtained and analyzed the broad band x-ray spectra for different orbital phases. The spectra from 3-70 keV obtained by LE, ME and HE detectors are analyzed here.

The continuum spectra of \gx have been studied extensively (e.g., F\"urst et al. 2018; La Barbera et al. 2005, and references therein). The X-ray spectrum of \gx at 3 --70 keV can generally be described with a partial covered power-law model, corrected with a high energy cutoff.  We implemented four kinds of phenomenological models to describe the continuum of \gx at first. Negative positive cutoff powerlaw or NPEX model (Mihara 1995), FDcut (Tanaka et al. 1986), HIGHECUT, newHcut (almost the same as highecut but being smoothed around the cutoff energy with a third-order polynomial; Burderi et al. 2000). We have smoothed the in-continuity of HIGHECUT with an additional Gabs component.

The models included two absorption components: one is the Galactic ISM absorption with the column density $NH_1$ being fixed at $NH_1=1.4\times 10^{22}$ cm$^{-2}$. Another column density parameter $NH_{2,pcf}$ is the partial covering photoelectric absorption in the binary system which is expected to vary with the orbital phases. We applied {\em tbnew}\footnote{http://pulsar.sternwarte.uni-erlangen.de/wilms/research/tbabs/} for both two components.

In the range of 6 --7 keV, there exist complicated emission features in \gx. These features are thought to be caused by fluorescence from neutral or ionized iron present in the material surrounding the neutron star. We added two Gaussian lines to fit the emission line features of \textit{Insight}-HXMT spectra: Fe K$\alpha$ (at $\sim 6.4$ keV) and Fe K$\beta$ ($\sim 7$ keV) lines. Three parameters for these emission features were summarized in Table~\ref{tab:specpar} including line centroid energy ($E_{Fe}$), line width ($\sigma_{Fe}$) and the equivalent width ($eqw_{Fe}$). The Fe K$\alpha$ line around 6.4 keV can be detected in all orbital phases, while the Fe K$\beta$ line around 7 keV could be detected by \textit{Insight}-HXMT only in part of the pointing observations (also see Figure~\ref{fig:multiexamspec}). In addition, as Ji et al. (2021) has confirmed, LE data is unable to constrain the line widths of the fluorescent lines. So, we fixed all these widths at 1 eV. Strong residuals at the lower energy side of Fe K$\alpha$ sometime appeared in our data and can be explained by an extended Compton shoulder (CS) (Watanabe et al. 2003; F\"urst et al. 2018; Ji et al. 2021). A box function was considered to describe this feature. However, in most of our observations, this feature is statistically insignificant and in principle, the energy resolution of LE telescope is not sufficient to fully resolve shape of the iron line complex. So, we did not fit this feature in the following analysis.

Absorption features around 30 --60 keV appeared in most of the observations, especially for those near the periastron passage. These features should be attributed to the cyclotron resonant scattering features (CRSF) of the magnetized accreting neutron star. We used a multiplicative absorption model with a Gaussian optical depth profile ({\em Gabs} in Xspec) to describe the cyclotron absorption line. In addition, a cross instrument calibration constant (CC) was added as the multiplicative components to calibrate the slight unequal flux between different telescopes of \hxmt. Thus, the baseline model in our analysis can be described as: 
 \be \rm CC*NH_1(NH_{2,pcf}*(CRSFs)*CONT+Gaussian).  \ee

With two {\em Gabs} components modeling the CRSFs, all of these models can obtain an acceptable fit in observations during the pre-periastron flare  (see Table~\ref{tab:fourmodel} and Figure~\ref{fig:multispecmodel}). We found that NPEX provided the best description of the overall spectrum with more stable continuum parameters (i.e. photon index and cutoff energy) and lower CRSF errors in most of the cases. In addition, the high energy part of physical model \emph{compmag} can be closely mimicked by the NPEX (F\"urst et al. 2018). We also found that the parameter values of CRSF and NH$_{2}$ have slight dependence on the continuum. In view of these premises and to compare our results with F\"urst et al. (2018), we continued our analysis of CRSFs with the NPEX model for all pointing observations. Spectrum parameters obtained in different orbital phases are summarized in Figure~\ref{fig:specresult}. Flux is calculated from the unabsorbed model in different energy ranges and hardness is evaluated from the ratio of flux in 8 --20 keV over the one in 5 --8 keV, i.e. Flux$_{8-20}$/Flux$_{5-8}$. We have tried several kinds of hardness ratio (10 --20/3 --10, 30 --70/3 --5 etc.) and found that the definition has only a small effect on its behavior during the orbital motion.

\begin{figure}
	\includegraphics[scale=0.33]{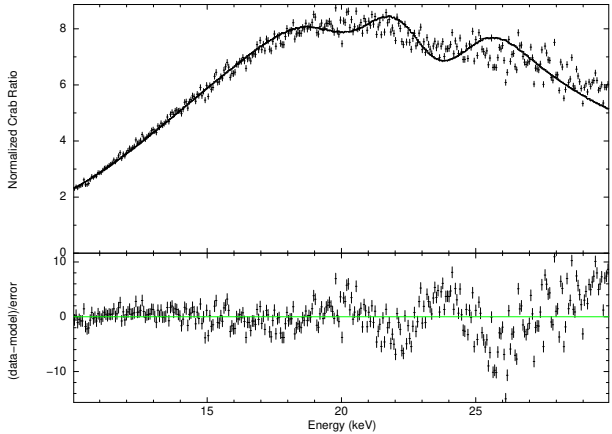}
	\caption{Crab ratio test: no feature detected in 10 --20 keV while an absorption feature at $\sim22$ keV was found, which can be fitted with a Gaussian absorption model. However, this narrow feature ($\sigma\sim1$ keV) is more likely to be an artifact due to the poor calibration in 21 --24 keV. We have ignored this energy range in our analysis.}
	\label{fig:ratiotest}
\end{figure}

\begin{figure*}
\includegraphics[scale=0.3]{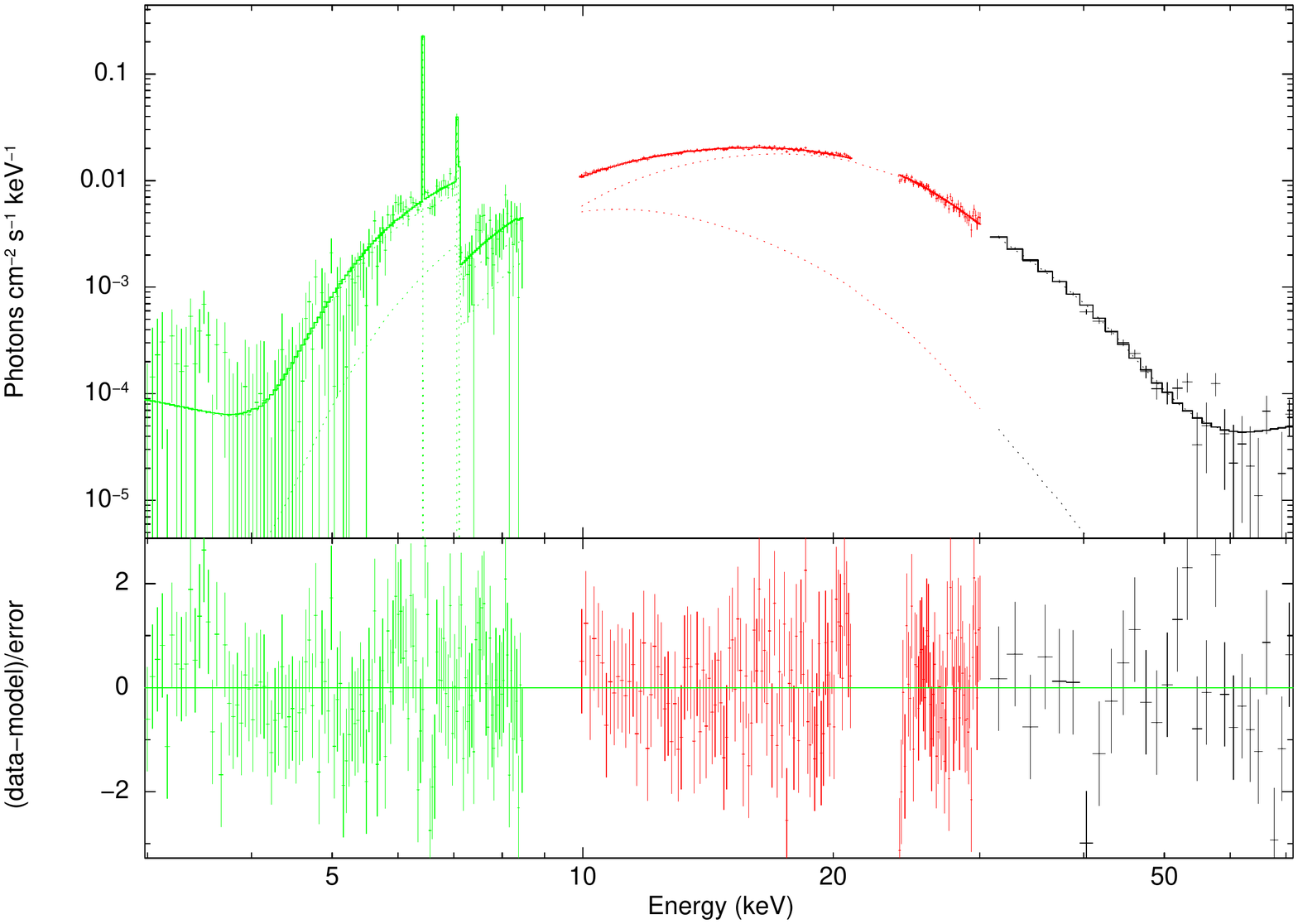}
\includegraphics[scale=0.3]{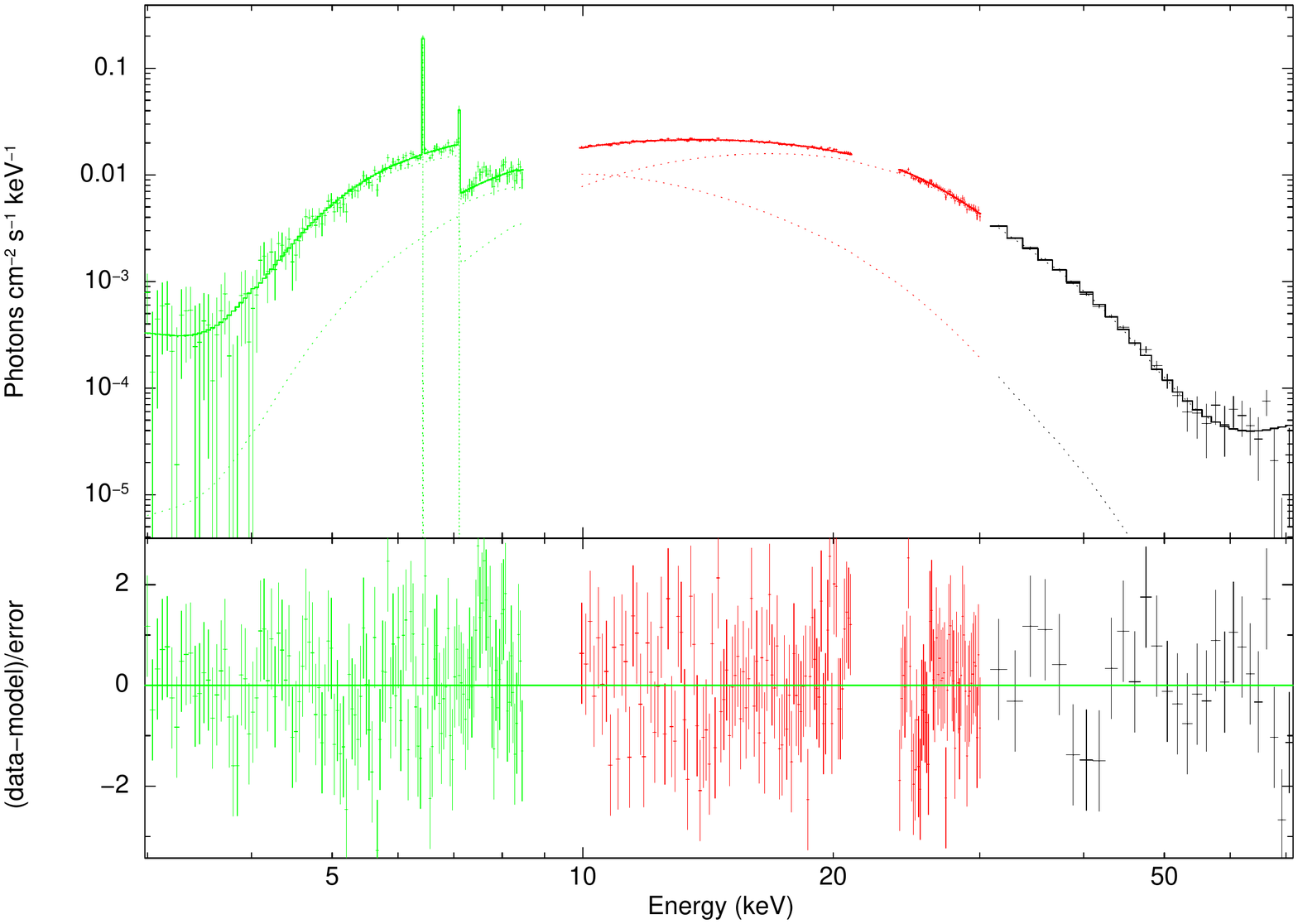}
\includegraphics[scale=0.3]{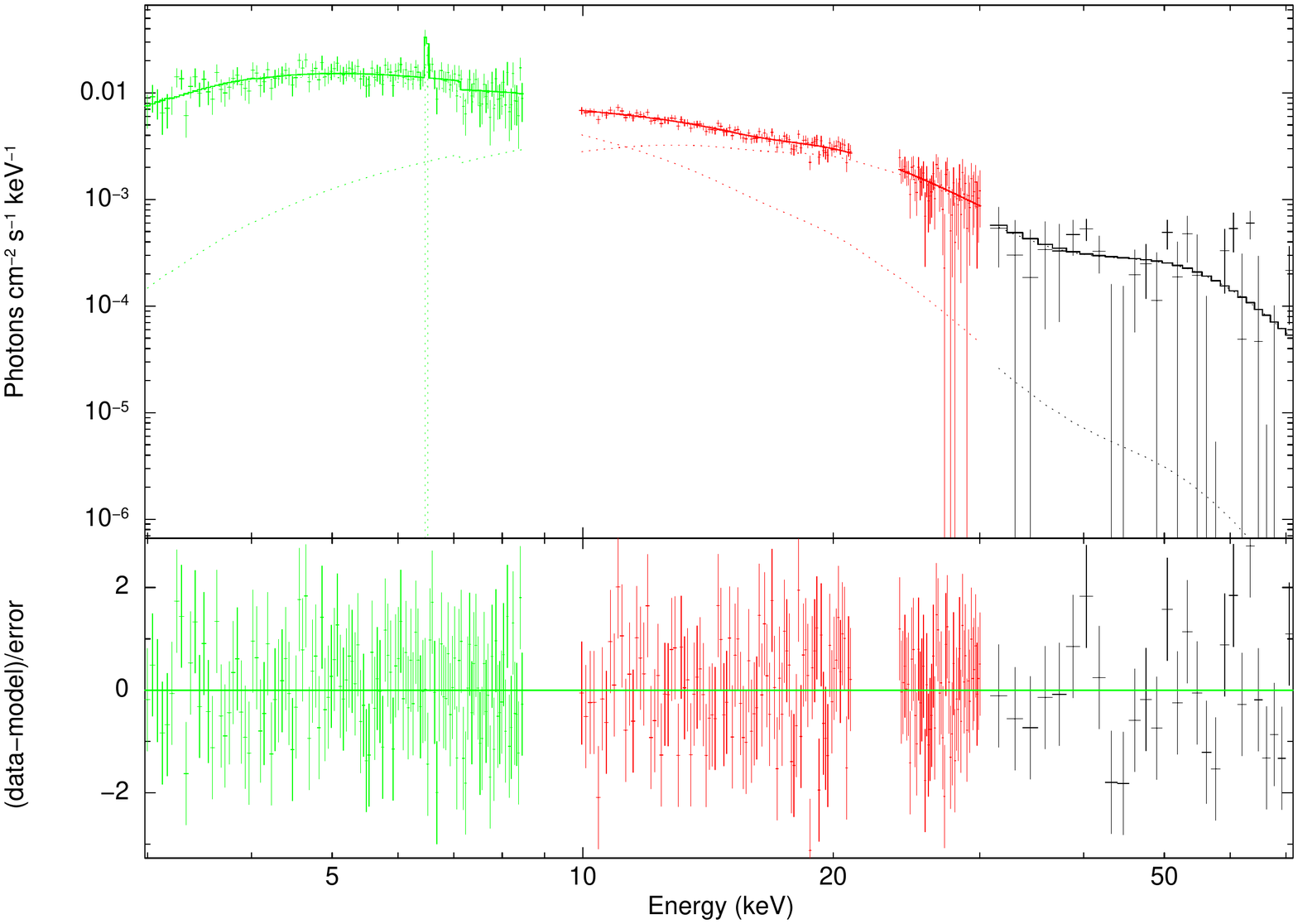}
\includegraphics[scale=0.3]{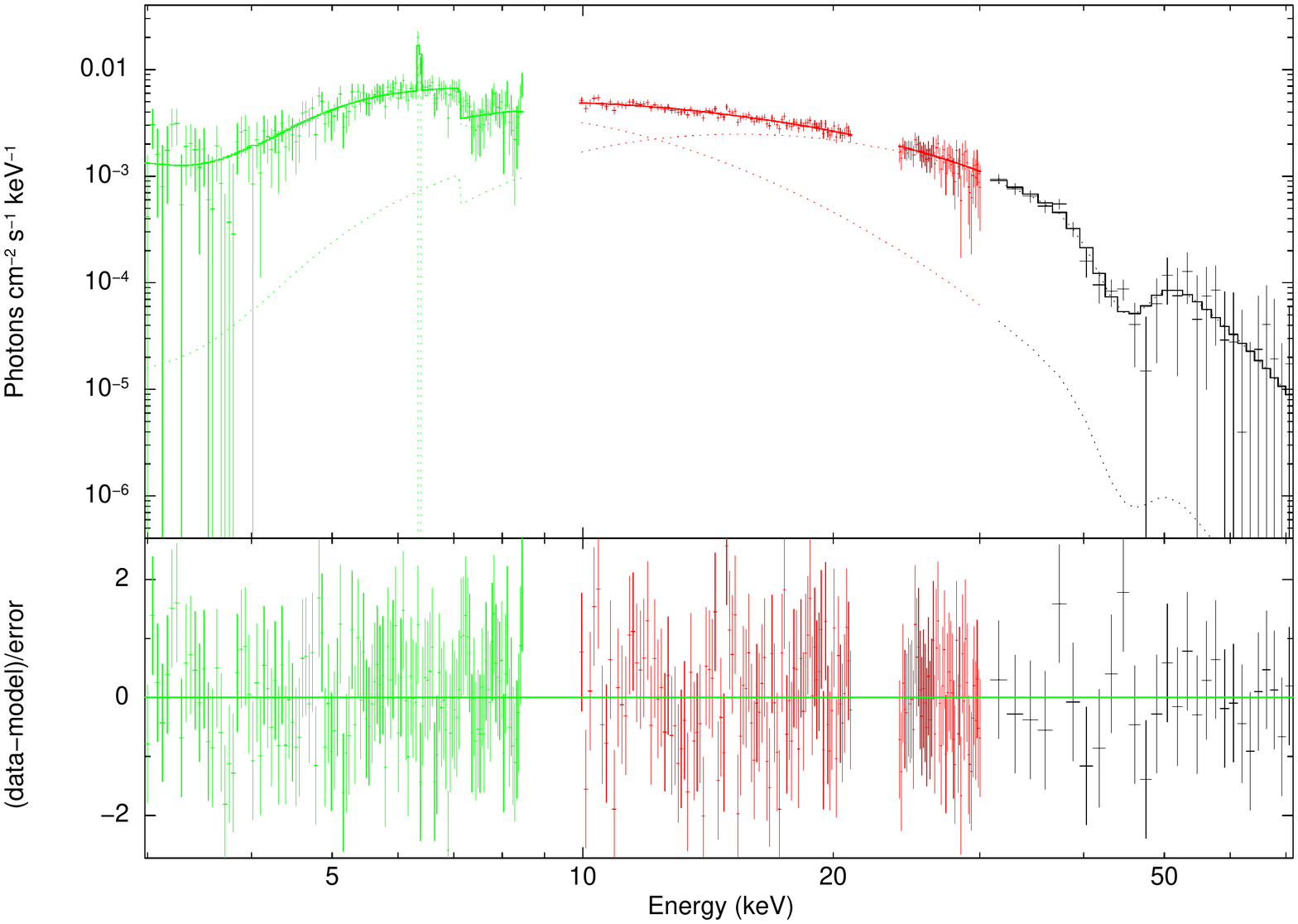}
\caption{Spectrum examples of \gx from 3--70 keV combined with the LE, ME and HE data. From the top to bottom, the panels presented the spectra with fitted model components for three different observations with ObsID: P010130900104 (phase=0.988) (top left); P010130900107 (phase=0.997) (top right); P010130901801 (phase=0.624) (bottom left); P020101228407 (phase=0.587) (bottom right). For observations near the periastron, we took two cyclotron absorption lines in fittings, at $\sim 30-40$ keV and $\sim 52-56$ keV while in the apastron phases, the S/N may be too low to distinguish/detect two absorption lines; in these cases, we took only one absorption line. The strength of the lower energy absorption is much smaller than the higher one near the periastron. Corresponding parameter values are in Table~\ref{tab:specpar}. }
\label{fig:multiexamspec}
\end{figure*}

\begin{figure*}
	\includegraphics*[scale=1]{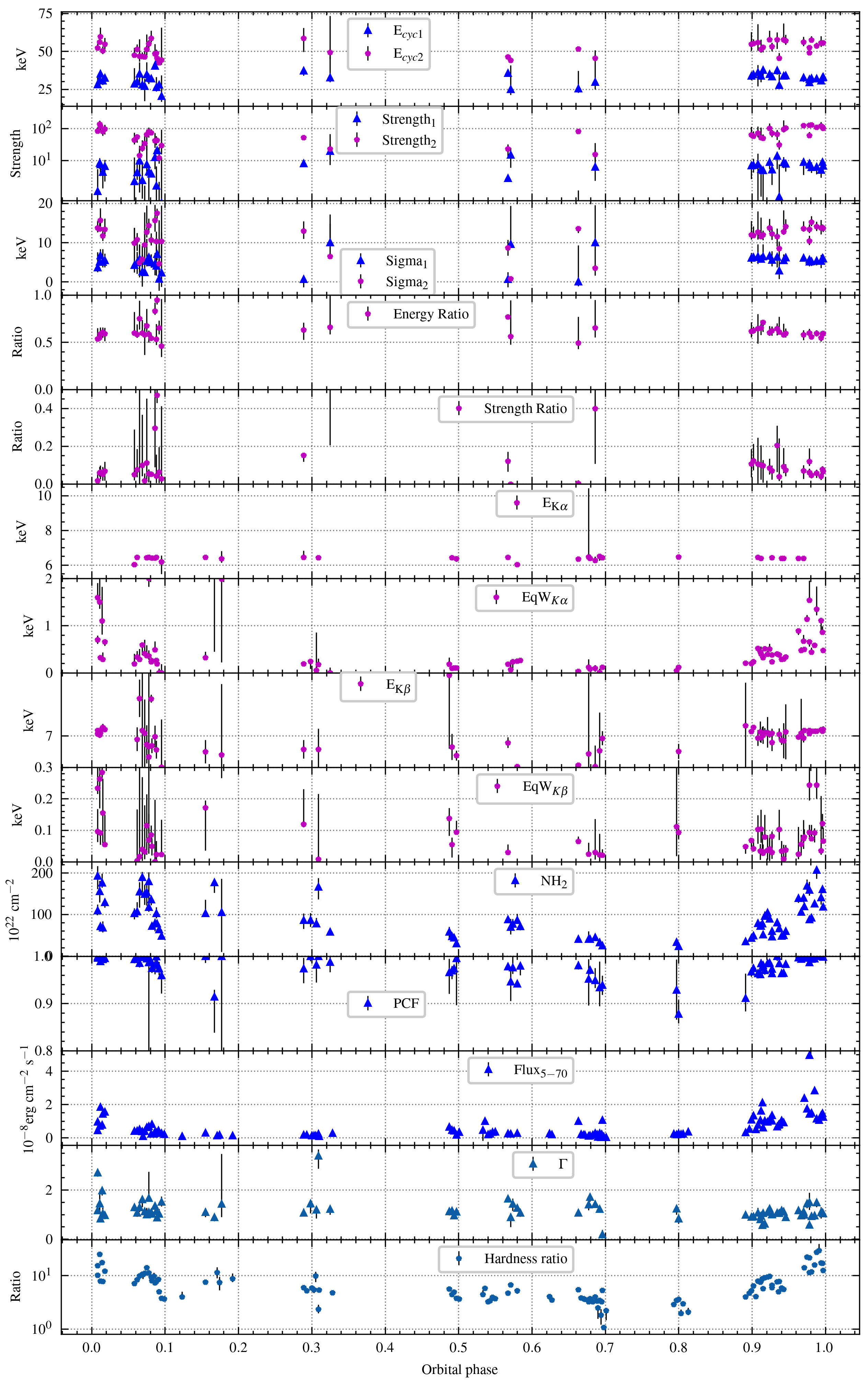}
\caption{Spectral parameters over the orbit of \gx. hardness is evaluated from the ratio Flux$_{8-20}$/Flux$_{5-8}$. NH$_2$ is the local column density in the binary system, showing peaks around the periastron orbital phases. PCF is the covering factor which is generally around 0.95 --1. The photon index is around 1.2 while $E_{\rm cut}\sim 7$ keV, showing no significant variations with orbit phases. The observed X-ray flux in the energy band of 5--70 keV in units of $10^{-9}$ erg cm$^{-2}$ s$^{-1}$ changes over the orbit, which correlates to the variation of NH$_2$. The equivalent width of Fe K$\alpha$ line around 6.4 keV also show variation over the orbit. Fe K$\beta$ line at $\sim 7$ keV was detected in some phases. In some observations, one or two CRSFs in $\sim$ 30--55 keV were reported.}
\label{fig:specresult}
\end{figure*}

We noticed wavy residuals between 10 and 30 keV in some observations and examined the possibility of the existence of another CRSF with the performance of Crab ratio test (see Figure~\ref{fig:ratiotest}). This ratio has the advantage of minimizing the effects due to the detector response and the uncertainties in the calibrations (Orlandini et al. 1998). We used observations that performed at the same month (i.e. April 2018), divided ME counts at each channel by Crab counts. Previously obtained continuum parameters were applied to fit this ratio while we alter the photon index to compensate for the $E^{2.1}$ factor. It was found that no prominent features appeared in 10--20 keV while a very narrow absorption feature was detected at $\sim 22$ keV. As we have already known that HXMT has large calibration uncertainty in 21--24 keV, this feature is more likely to be instrumental.

In Figure~\ref{fig:multiexamspec}, we showed the spectra from 3--70 keV obtained by three detectors LE, ME and HE for different orbital phases. In some observations, after fitting the spectrum with the cyclotron absorption line around 52--55 keV, there still exist an absorption feature in lower energies around 33 keV). We confirmed that this feature can be fitted with another Gaussian absorption feature. We adopt Monte Carlo Markov Chain (MCMC) method to estimate the error and detect any parameter degeneration. During the primary run, each chain was generated using 200000 steps (20 walkers) with the first 10000 steps being burnt. We found that in different orbital phases, the energy ratio of these two lines are fluctuating around a constant while the corresponding MCMC contour illustrate no apparent degeneration between these two parameters (see Figure~\ref{fig:MCMC}). Thus, it is reasonable to claim that the correlations are intrinsic of the system.

\begin{table*}
	%\tabletypesize{\scriptsize}
	\scriptsize
	\caption{Spectral parameters of \gx fitted with NPEX based on some \textit{Insight}-HXMT observations. X-ray flux is derived from 3--70 keV, while the hardness ratio is evaluated from the ratio of Flux$_{8-20}$/Flux$_{5-8}$.}
	\label{tab:specpar}
	\renewcommand\arraystretch{1.8}
	\renewcommand\tabcolsep{3.0pt}
	% \setlength{\tabcolsep}{1.0mm}
	%\tablewidth{0pt}
	\begin{center}
		\begin{tabular}{l l l l l }
			\hline \hline
			Observations & P010130900104&P010130900107&P010130901801&P020101228407\\
			\hline
			NH$_2$ ($10^{22}$ cm$^{-2}$) &	$206^{+9}_{-20}$&	$118^{+6}_{-5}$ &$19^{+7}_{-3}$ &	$77^{+13}_{-11}$\\
			PCF   &	$0.994^{+0.005}_{-0.025}$&	$1.000_{-0.049}$ &	$0.95_{-0.062}^{+0.031}$&	$0.974^{+0.0012}_{-0.012}$\\
			$\Gamma$  &	$1.492^{+0.054}_{-0.292}$& $1.02^{+0.06}_{-0.13}$&	$0.771^{+0.16}_{-0.09}$ &	$1.15^{+0.25}_{-0.16}$\\
			E$_{\rm cut}$ (keV) &	$7.096^{+0.437}_{-0.256}$&	$7.24^{+0.10}_{-0.27}$&	$6.58^{+0.06}_{-0.14}$	& $6.28^{+0.71}_{-0.39}$\\
			E$_{\rm Fe K\alpha}$ (keV) & $6.425^{+0.003}_{-0.004}$& $6.434^{+0.001}_{-0.002}$& $6.49_{-0.27}^{+0.81}$ &	$6.36^{+0.07}_{-0.16}$\\
			EqW$_{\rm Fe K\alpha}$ (keV)&	$1.68^{+0.15}_{-0.15}$&	$0.492^{+0.044}_{-0.005}$  & $0.080^{+0.067}_{-0.04}$	 & $0.16_{-0.12}^{+0.03}$  \\
            E$_{\rm Fe K\beta}$ (keV) &$7.06^{+0.02}_{-0.01}$ &$7.09^{+0.02}_{-0.01}$& - & - \\
            EqW$_{\rm Fe K\beta}$ (keV)& 	$0.25^{+0.12}_{-0.01}$ &$0.080^{+0.007}_{-0.003}$& - & - \\
			E$_{\rm cyc1}$ (keV) & $31.8^{+0.7}_{-0.9}$	 &$33.8^{+0.7}_{-0.4}$&	$29^{+5}_{-2}$   &$44^{+25}_{-1}$ \\
			$\sigma_{\rm cyc1}$ (keV) & $5.3^{+0.5}_{-0.4}$&$6.2^{+0.6}_{-1.0}$& $11.9^{+0.4}_{-6.2}$	&$3^{+14}_{-2}$\\
			Strength1 &	$5^{+2}_{-2}$ &$7.3^{+2.5}_{-1.8}$&	 $19^{+20}_{-10}$	& $11^{+90}_{-1}$\\
			E$_{\rm cyc2}$ (keV) &	$53.5^{+2.0}_{-1.2}$&	$56.24^{+0.6}_{0.7}$ & -	&	- \\
			$\sigma_{\rm cyc2}$ (keV) &$14.5^{+1.9}_{-0.4}$&	$14.3^{+0.3}_{-1.0}$&	- &	-\\
			Strength2 &	$115^{+30}_{-12}$ &	$105^{+4}_{-14}$ & - &-\\
		    Flux $10^{-8}$ erg s$^{-1}$ cm$^{-2}$ &  $1.133^{+0.005}_{-0.008}$  &	$1.272^{+0.005}_{-0.004}$ & $0.19^{+0.02}_{-0.02}$ &	$0.268^{+0.009}_{-0.048}$ \\
		    Hardness Ratio &  $26^{+2}_{-3}$  &	$10.2^{+0.3}_{-0.4}$ & $2.9^{+0.4}_{-0.2}$ &	$3.9^{+0.7}_{-0.3}$ \\
			Reduced-$\chi^2$ (dof) & 1.1064 (287) &	1.1353 (287)& 0.9266 (292)	 & 0.7263 (292)\\
           	\hline \hline
		\end{tabular}
	\end{center}
\end{table*}

We applied the NPEX with one or two absorption lines to fit all observed spectra from 3--70 keV based on \textit{Insight}-HXMT observations. In the apastron phases, the S/N may be too low to distinguish the second line. In most situations, we took two cyclotron absorption lines in fittings. For some observations near the apastron, the S/N may be too low to distinguish two absorption lines (these features having very small strength), we took only one absorption line then. The model has given good fit to these spectra, with the reduced $\chi^2$ around 1 (see Table~\ref{tab:specpar}). All spectral parameters and their variations over the orbit are then collected and plotted in Figure~\ref{fig:specresult}. The spectral parameters really had some changes with orbital phases. X-ray flux and the intrinsic column density show nearly simultaneous variations over orbit, both of which have peaks around pre-periastron and post-periastron orbital phases as expected. The covering factor did not change within a narrow range $\sim 0.95-1$. The cutoff energy $E_{\rm cut}\sim 7$ keV and photon index $\Gamma\sim 1$ did not show significant variation over orbit.

In all orbital phases, the Fe K$\alpha$ line around 6.4 keV was detected, with the equivalent width evolving with the orbital phases. The iron line properties may be related to the column density in the accretion system, which will be discussed in \S 5.1. The cyclotron absorption lines were detected in the X-ray spectra of \gx, which are at $48-56$ keV and sometimes, another line at $\sim 28-36$ keV was also reported. In addition, line energies of these two CRSFs have some correlations. We will discuss it in \S 5.2.

\begin{figure}
	\includegraphics[scale=1]{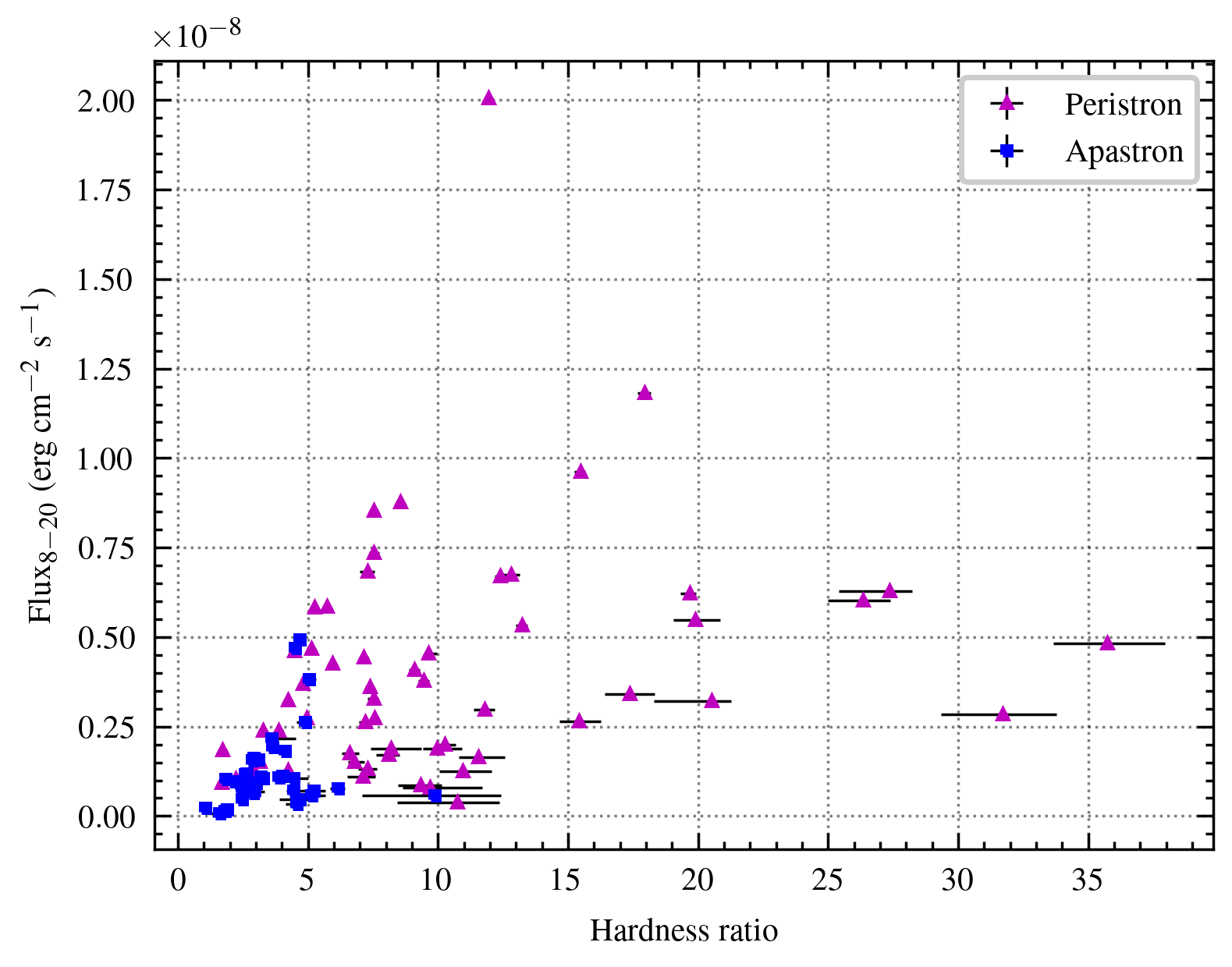}
	\caption{Hardness-Intensity diagram of \gx. Hardness ratio is generally positive correlated with source flux. }
	\label{fig:hid}
\end{figure}

\section{Discussions}

\begin{figure*}
\includegraphics[scale=1]{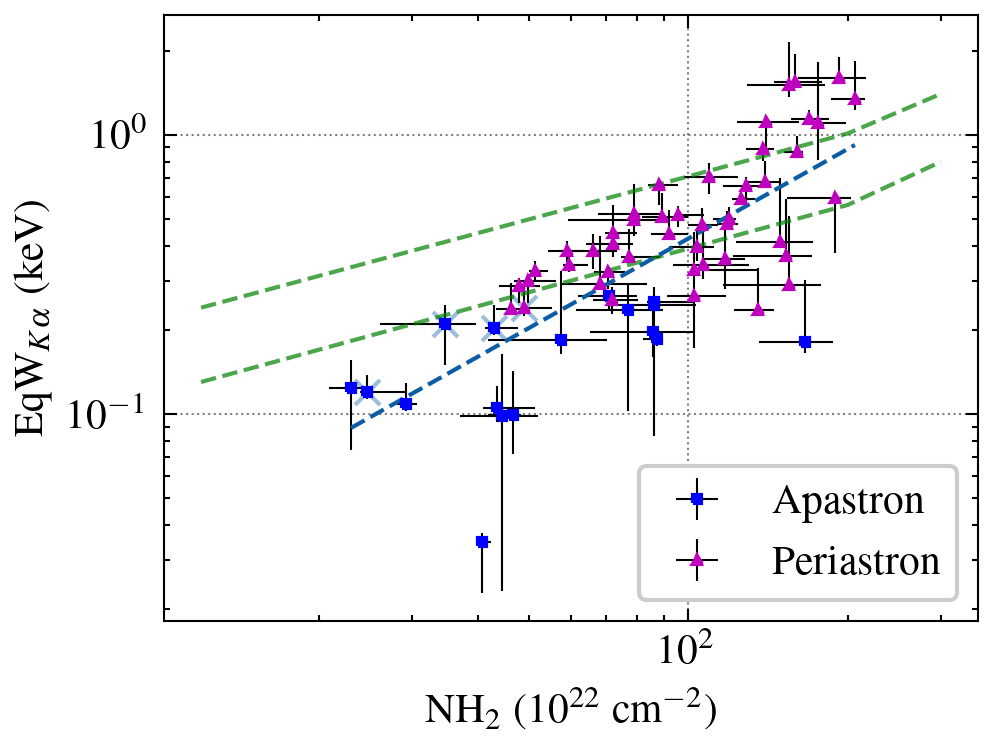}
\includegraphics[scale=1]{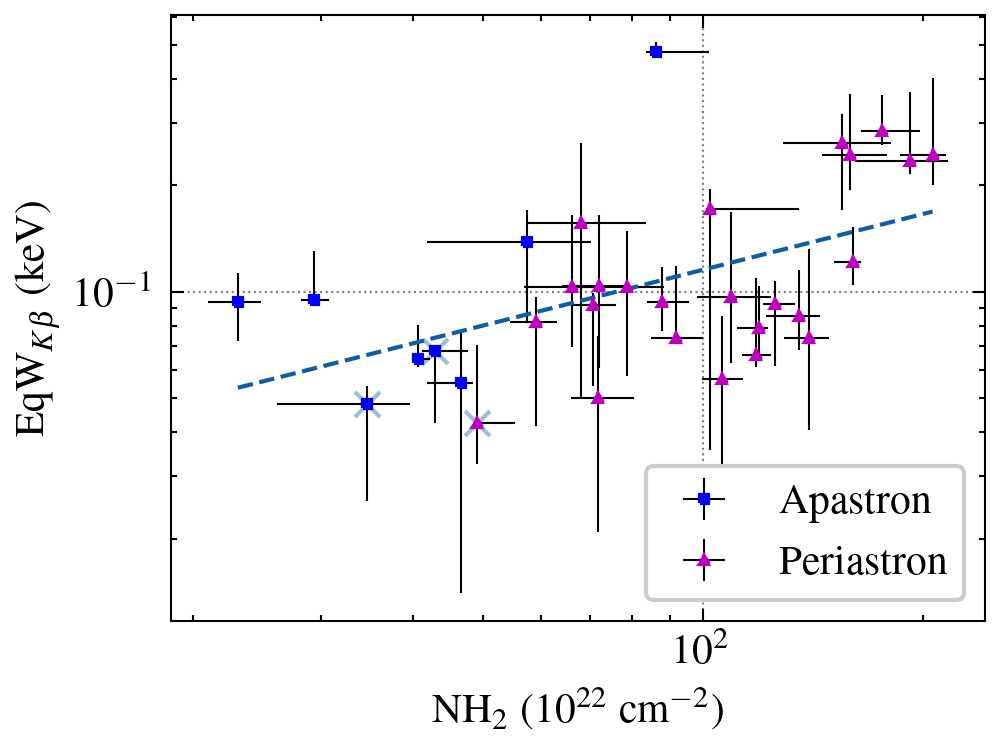}
\caption{The equivalent width versus the hydrogen column density for two iron lines of \gx: Fe K$\alpha$ line at $\sim 6.4$ keV (left) and Fe K$\beta$ line at $\sim 7$ keV (right). The equivalent width values of both two iron lines show positive correlation with the column density. Blue dashed lines represent the best fitted broken powerlaw functions. Data points being taken from the assumed disc accretion stage (MJD 58400--58600) were marked with blue crosses. Green dashed lines are the EqW-NH$_{2}$ relation derived by numerical simulation (assuming spherical symmetric, from Gimenez-Garcia et al. 2015), with the corresponding photon indexes of 0.5 and 2.0 respectively.
}
\label{fig:cog}
\end{figure*}
\begin{figure*}
	\includegraphics[width=\columnwidth]{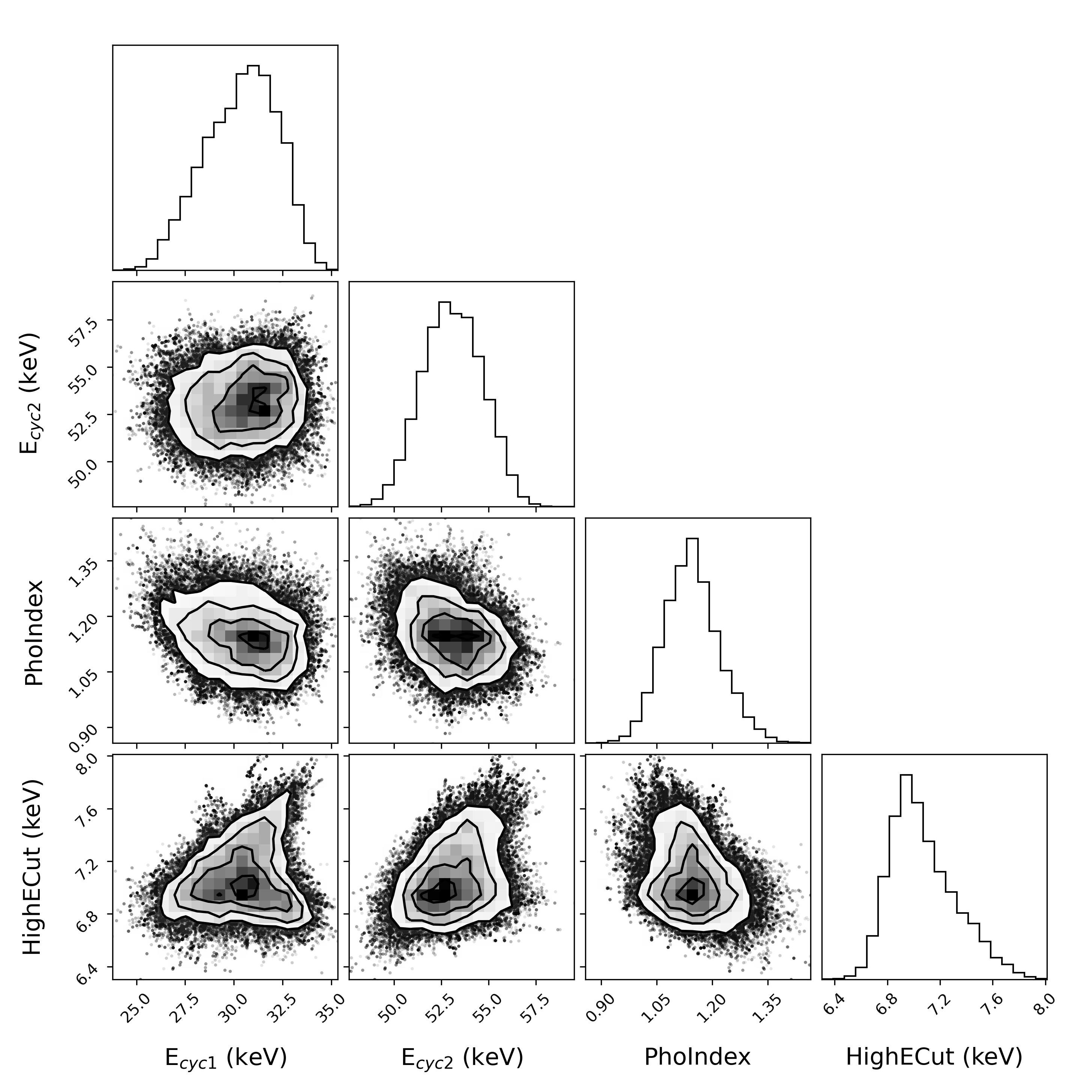}
	\caption{Probability distribution of CRSF and continuum parameters during fitting process generated by MCMC. There is no apparent parameter degeneration between CRSF centroid energies. In addition, there is only a very weak correlation between continuum parameters and CRSF energy.}
	\label{fig:MCMC}
\end{figure*}

\begin{figure*}
	\includegraphics[scale=1]{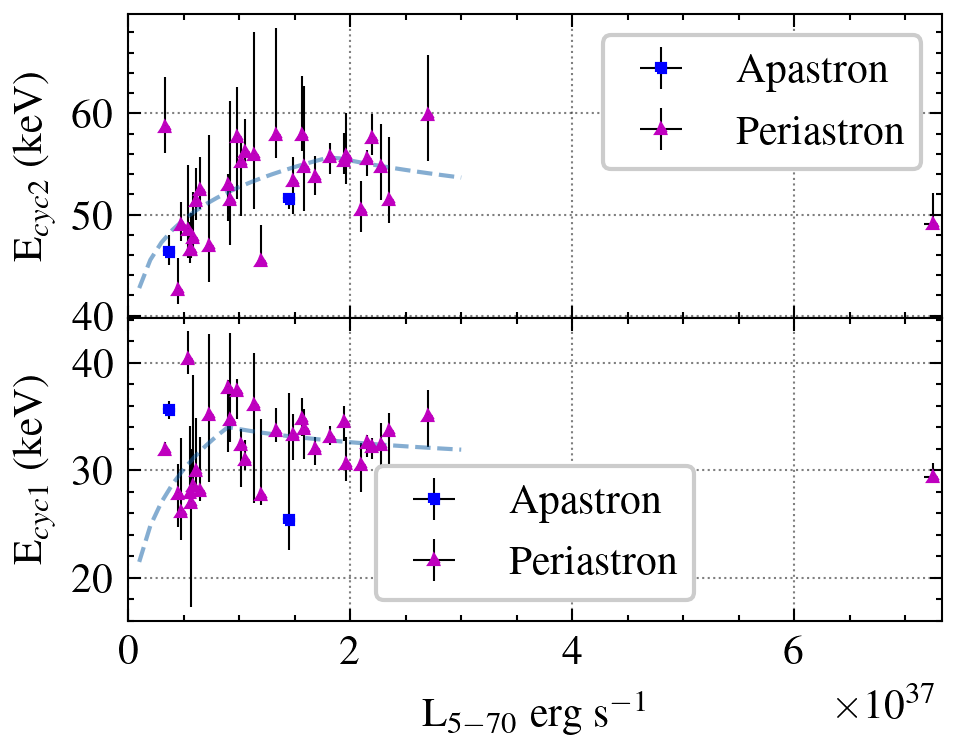}
	\includegraphics[scale=1]{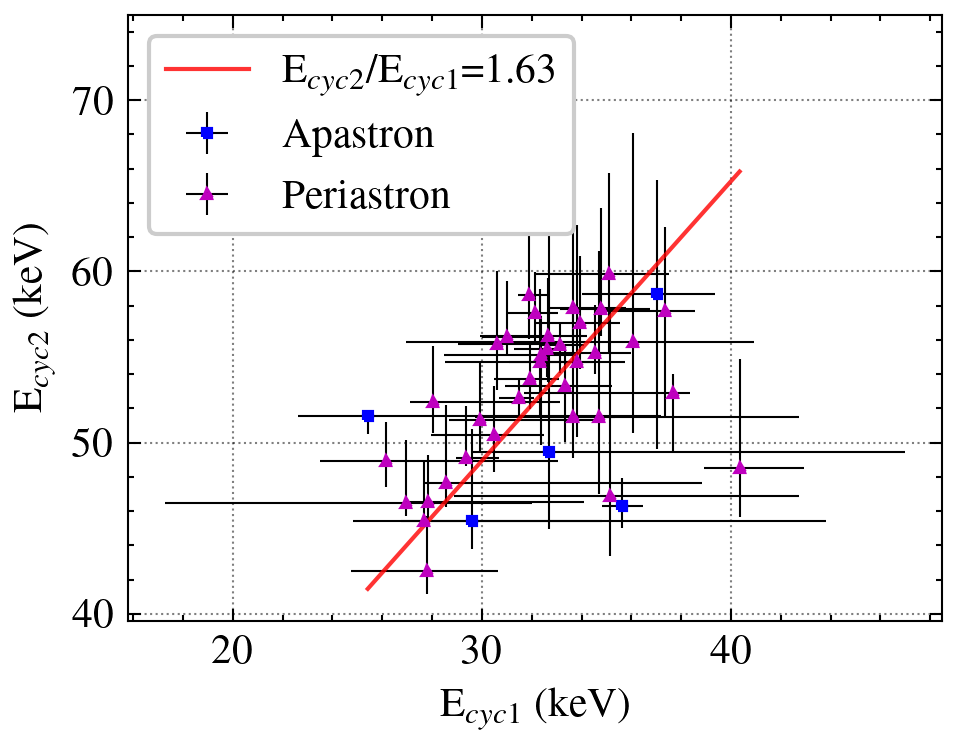}

	\caption{{\bf Left panel:} { Cyclotron absorption energy dependence of source unabsorbed luminosity in 5 --70 keV. Positive correlation between CRSF energy and X-ray luminosity was observed in the low luminosity state, while above a critical luminosity around $10^{37}$ erg s$^{-1}$, the CRSFs energy correlate with the luminosity negatively. Blue dashed lines are broken powerlaw fittings, producing a best-fitting critical luminosity.} {\bf Right panel:} CRSF line energy correlation of \gx. Two CRSFs have strong linear correlation especially at pre-periastron flare, which is not likely to be the consequence of parameter degeneration (also see Figure~\ref{fig:MCMC}).
	}
	\label{fig:cyccor}
\end{figure*}

\subsection{Curve of growth}

Theoretically, it is expected that the strength of spectral lines are positively correlated with the absorber optical depth. The equivalent width of Fe K$\alpha$ line shows positive correlation with $NH_2$ which has been generally known as the curve of growth (see Torrejon et al. 2010; Gimenez-Garcia et al. 2015). Here we can probe the relation between the column density and iron lines observed in \gx. In \S 4, we have obtained spectral properties over the orbit including two iron lines at 6.4 keV and 7 keV, their equivalent widths, and the corresponding intrinsic absorption column density NH$_2$ for each observation. $NH_2$ is around the level of $10^{24}$ cm$^{-2}$ at periastron orbital phases, and $NH_2\sim (1-5)\times 10^{23}$ cm$^{-2}$ at other phases. In Figure~\ref{fig:cog}, we present the hydrogen column density $NH_2$ versus the equivalent width (EqW) of both two iron lines: Fe K$\alpha$ line and Fe K$\beta$ line. 

For the case of \gx, we confirmed a strong correlation, and a power-law fit in Figure~\ref{fig:cog} leads to \be eqw \propto NH_2^{1.06\pm 0.20}, \ee with the Pearson coefficient (PC) of 0.75.

In addition, we also obtained the equivalent width of the Fe K$\beta$ line in some observations of \textit{Insight}-HXMT near the pre-periastron orbital phases. Thus we also plotted the $eqw$ of Fe K$\beta$ versus $NH_2$ in Figure~\ref{fig:cog}. We found the curve of growth of the Fe K$\beta$ line in \gx. The correlation of $eqw_{Fe K\beta}$ versus $NH_2$ is not so strong, the powerlaw fit in Figure~\ref{fig:cog} leads to \be eqw \propto NH_2^{0.52\pm 0.28}, \ee with the Pearson coefficient of 0.51.

The derived correlation between the equivalent width of Fe lines and intrinsic column density is not completely consistent with the early work by Ji et al. (2021). This is due to the larger sample and broader energy band (3--70 keV compared to 5.5--8.5 keV) we have adopted. In other high mass X-ray binaries, this relation was also reported (Torrejon et al. 2010; Gimenez-Garcia et al. 2015). \gx shows the similar behavior to that of other X-ray binaries (see the good example of IGR J16320-4751 in Gimenez-Garcia et al. 2015). From the left panel of Figure~\ref{fig:cog}, $eqw_{Fe K\alpha}\sim 0.1$ at $NH_2=3\times 10^{23}$ cm$^{-2}$, increased to $eqw_{Fe K\alpha}\sim 1$ at $NH_2=2\times 10^{24}$ cm$^{-2}$. During the periastron flare, the curve of growth for \gx is in good agreement with the numerical simulation shown in Gimenez-Garcia et al. (2015). However, at apastron, our result showed a significant deviation from the spherical accretion scenario (EqW values are lower than the simulations, see Figure~\ref{fig:cog}).

\subsection{Cyclotron absorption line energies}

\gx has been studied extensively, and the cyclotron absorption line energy varied in a wide range from $\sim 30 -56$ keV based on previous observations. \textit{Insight}-HXMT detected the cyclotron lines both in periastron and apastron orbital phases, where two cyclotron absorption lines with a Gaussian profile are reported, one line around $\sim 26-35$ keV, and the other around $\sim 45-56$ keV. Two possible absorption lines at $\sim 35-40$ keV and $\sim 50-55$ keV reported by NuSTAR (F\"urst et al. 2018) were then confirmed by \textit{Insight}-HXMT. This interpretation can explain the wide energy range and wild variability of the line parameters in previous study.

Combining about 40 pointing observation data, we analyze the CRSF line parameters systematically. From Figure~\ref{fig:cyccor}, we found that the two centroid energies of CRSFs have a very strong linear correlation, indicating a fixed ratio between them. We confirmed the result is not due to the parameter degeneration during fitting process (see Figure~\ref{fig:MCMC}). This ratio do not scale like 1 :  2 but is about 1 : 1.63, which is significantly greater than that derived by F\"urst et al. (2018) but still too small compared to the value predicted by M\'esz\'aros (1992). It is suggested that there may exist two line forming regions (F\"urst et al. 2018), situated at the surface of NS and 1.4 km above it respectively. This interpretation is based on the premise of Coulomb-radiation-dominated deceleration regime. The corresponding shock height can be calculated from Eq. (51) in Becker et al. (2012), assuming radius of the NS to be 10 km and a mass of 1.8$M_{\sun}$, we have:
	\begin{equation}
		h=2.2\times 10^{4}L_{37}^{-5/7}B_{12}^{-4/7}\Lambda^{-1}\; \rm cm.
		\label{eqn:subcriheight}
	\end{equation}

As the shock height should decrease with increasing luminosity, this regime will require a positive correlation of the CRSF energy with luminosity. This correlation was observed in low luminosity stage for both two CRSFs simultaneously (see left panel of Figure~\ref{fig:cyccor}). Two forming region interpretation (F\"urst et al. 2018) claimed that the higher energy absorption line around 50 keV was situated at the surface of NS, thus being unable to account for the strong correlation between the line energy and X-ray luminosity, unless the surface magnetic field of the neutron star also evolves during the orbital motion (should be impossible). 

In addition, in more luminous range, we observed a negative correlation of the line centroid energy and X-ray luminosity for both two lines, which suggested the existence of a critical luminosity in \gx. We have taken a broken powerlaw model to fit the relation of $E_{cyc} - L_x$ for both two lines. The cutoff luminosity is determined to be $(1.83\pm 0.46)\times 10^{37}$ erg s$^{-1}$ for the high energy line, and $(8.99\pm2.53)\times 10^{36}$ erg s$^{-1}$ for the lower one (see left panel of Figure~\ref{fig:cyccor} for more details on the position of the determined critical luminosity in the fittings). Thus, we at first derived the critical luminosity in \gx from the X-ray observations. At critical luminosity, the radiation pressure at the base of neutron star accretion column can contribute significantly to the deceleration of infalling materials, which can defined as the function of the cyclotron absorption line energy (Becker et al. 2012): 
\begin{equation}
\begin{split}
L_{\rm crit}\sim & 1.49\times 10^{37} {\rm erg\ s^{-1}} \Lambda_{0.1}^{-7/5} \\
&\times (\frac{M_{\star}}{1.4M_{\odot}})^{29/30} (\frac{R_{\star}}{10\;\rm km})^{1/10} ({E_{cyc}\over 11.6 \rm keV})^{16/15},
\end{split}
\label{eqn:criticallumi}
\end{equation}
where $M_{\star}$ and $R_{\star}$ are the mass and radius of neutron star, respectively; $\Lambda=1$ in the case of spherical accretion, and $\Lambda\sim0.22\alpha^{18/69}$ in disc accretion case (Becker et al. 2012; Harding et al. 1984).

Considering the uncertainty, we suggested the observed critical luminosity range in \gx of $(0.7-2.2)\times 10^{37} \rm erg\ s^{-1}$. Although during the spin-up event (MJD 58400-58600, only four available observations) \gx was considered having an accretion disc, observations in this stage still had only very low luminosity and thus low S/N for spectral analysis. Consequently, none of them were plotted on Figure~\ref{fig:cyccor}.  In a word, wind accretion may dominate the accretion process in our data, which generally should be a spherical symmetric process. According to Equation~\ref{eqn:criticallumi}, given $\Lambda=1$ for spherical symmetric accretion, we found that the energy of the CRSF on the surface of neutron star reaches the range of $\sim 90 - 200$ keV, suggesting a surface magnetic field of $(8-20)\times 10^{12}$ G in \gx. Thus the neutron star in \gx would be a strongly magnetized neutron star of typical magnetic field in the order of $10^{13}$ G (a little lower than the typical magnetar-like field suggested by Doroshenko et al. 2010). We concluded that the observed cyclotron absorption lines at energies of 30 -- 56 keV are likely to be produced at the height $>10^5$ cm above the star surface, supporting the existence of a tall accretion column structure during periastron flare.

\begin{figure}
	\includegraphics[width=\columnwidth]{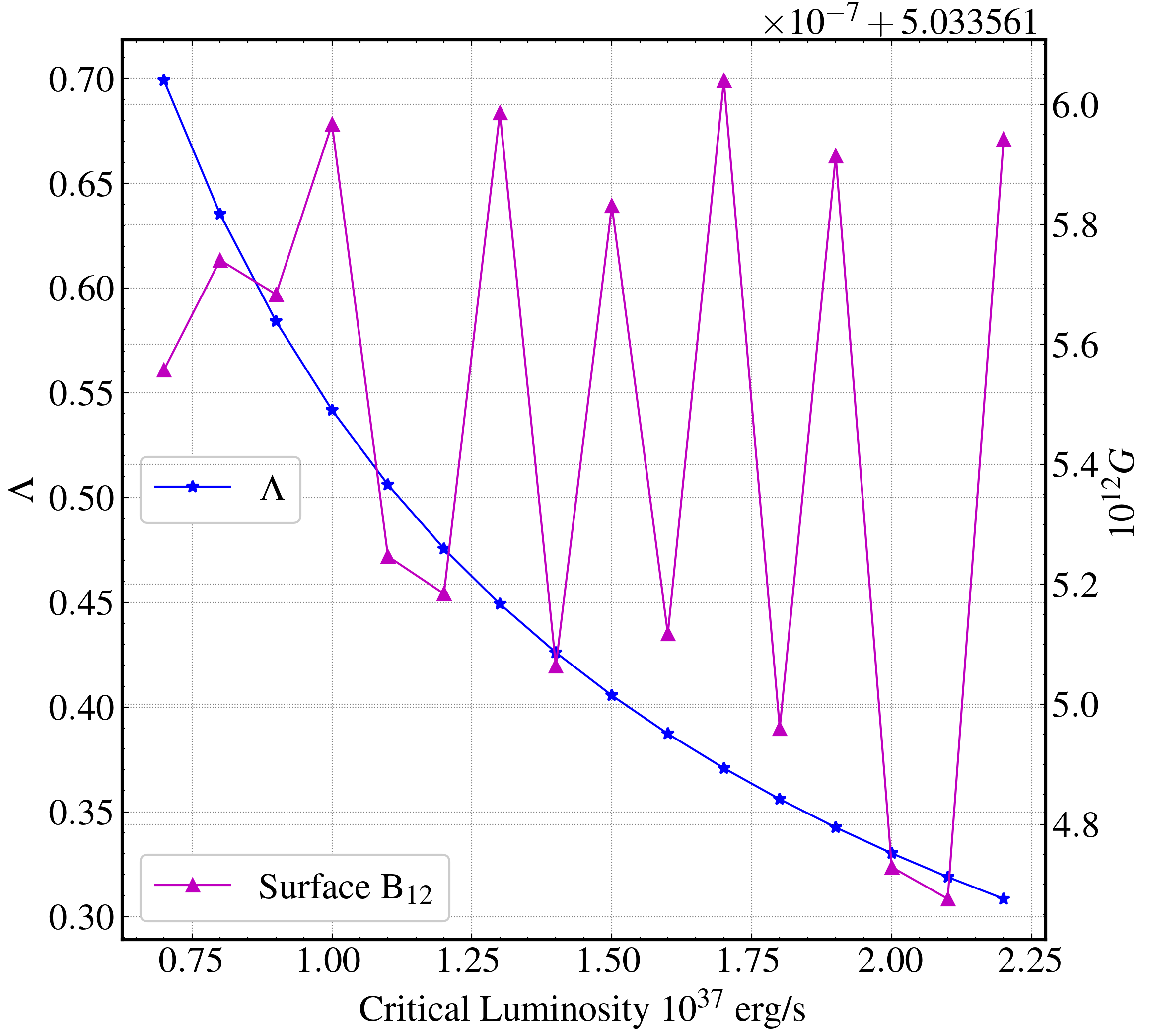}

	\caption{Calculated relation of $\Lambda$-critical luminosity and surface magnetic field strength-critical luminosity.}
	\label{fig:lambdatheo}
\end{figure}

However, such a CRSF emission area height actually contradicts with Equation~\ref{eqn:subcriheight}, which indicates an emission height tenfold lower than what we have derived by the critical luminosity. This may make our interpretation self-contradicting. Here, we proposed another possibility that the parameter $\Lambda$ may not be 1, because the result in Figure~\ref{fig:cog} has shown significant deviation from isotropic. However, there is no any conclusive evidence that could prove the existence of disc accretion either. Consequently, assuming a dipole magnetic field structure, we did a numerical calculation to obtain possible $\Lambda$, keeping the emission height derived by Becker et al. (2012) equal to the observed values (critical luminosity and CRSF centroid energy are derived from the broken powerlaw fitting in Figure~\ref{fig:cyccor}), the corresponding results are shown in Figure~\ref{fig:lambdatheo}. The $\Lambda$ values are generally less than 1 (wind accretion) but greater than 0.1 (disc accretion) while the best-fitting surface magnetic field stayed at $\sim 5\times10^{12}$ G. This result implies a complicated accretion environment including a blending of wind and disc accretion.

We also compared our results with the model proposed by Mushtukov et al. (2015). Assuming $\Lambda=0.5$ and equal flux for X- and O-modes, the calculated critical luminosity for a NS with magnetic field of $\sim 5\times10^{12}$G is $\sim0.8-2.0\times10^{37}$erg s$^{-1}$, corresponding to $l_{0}/l$=0.5 and 1.0 respectively (see Figure 7 of their paper). This result is also in good agreement with our observation. Considered the large uncertainty in $\Lambda$ and mixing polarization modes, the difference between their model and the one from Becker et al. (2012) may not be significant.

We noticed that in most of the orbital phases, especially that near the periastron (see Figure~\ref{fig:specresult}), the strength ratio of these CRSFs fluctuates around $\sim$ 0.05. This result indicates that we may observe only one line but with a significant deviation from a simple empirical function, which is due to the superposition of CRSFs in different forming regions with different heights (Nishimura 2015). Such deviations have also been theoretically predicted by Schwarm et al. (2017) with MC simulations. Future numerical simulations and more detailed comparison between observations and simulated results could probe the physical origin of the very broad absorption feature.

\section{Summary and Conclusion}

\gx is an interesting X-ray pulsars showing significant variation with the orbital phases both in light curves and spectral properties. \textit{Insight}-HXMT has performed multiple pointing observations on this source covering different orbital phases, and found the timing and spectral variations over orbit in \gx. The spin period of \gx showed variations, in addition, from 2017-2020, the neutron star of \gx undergone a long-term spin-up state with the period evolving from $\sim 685$ s to 670 s. The pulse profiles changed in three energy bands and also varied with orbital phases. Pulse fractions from the bands of 3- 60 keV showed negative correlation with the X-ray flux.

The spectral parameters evolved over orbit in \gx. The X-ray flux correlates to the column density, showing peaks near the periastron and apastron orbital phases. The Fe K$\alpha$ line was detected in all orbital phases. In some orbital phases, the Fe K$\beta$ line at $\sim 7$ keV was also detected. The curve of growth for both two iron fluorescence lines were obtained, their equivalent width have a positive correlation with the column density.

Two CRSFs at $\sim 30-42$ keV and $\sim 50-56$ keV in \gx were confirmed by the \textit{Insight}-HXMT observations. The cyclotron line energies showed slight variation over orbit and strong correlation with each other, having a fixed ratio $\sim1.63\pm0.01$. The strength ratio between these two features stayed at $\sim 0.05$ during most of the phases.  A simultaneous increase of line energy with X-ray luminosity for both two CRSFs was observed in low flux range, while in luminous states, both the two line energies showed a negative relation with the luminosity. The observed critical luminosity is around $10^{37}$ erg s$^{-1}$. Thus, if wind accretion does dominate \gx, we estimated the surface magnetic field of the neutron star in \gx as $(1-2)\times 10^{13}$ G. Another possibility is that \gx undergoes a blending of disc and wind accretion during periastron flare, with a magnetic field of $\sim 5\times 10^{12}$ G. The environment there is more complicated than we have postulated. In any case, \gx should be a highly magnetized neutron star, but its field is still lower than the typical field of magnetars (Doroshenko et al. 2010; Wang 2013). The observational phenomena indicated a strong coupling between two CRSFs. The two forming region interpretation proposed by F\"urst et al. (2018) could not account for such a correlation between the line energies and luminosity. We proposed that we observed only one line but with a significant deviation from a simple empirical function. According to the observed critical luminosity, if wind accretion predominate, the line forming region may be high above the neutron star surface. The future simulated physical CRSF profile is expected to account for the strange line profile in \gx.

\section*{Acknowledgments}
We are grateful to the referee for the fruitful and constructive suggestions and comments to improve the manuscript. The work is supported by the NSFC (U1838103, 11622326, U1838201, U1838202), the National Program on Key Research and Development Project (Grants No. 2016YFA0400803, 2016YFA0400800). This work made use of data from the \textit{Insight}-HXMT mission, a project funded by China National Space Administration (CNSA) and the Chinese Academy of Sciences (CAS).

\section*{Data Availability}

Data that were used in this paper are from Institute of High Energy Physics Chinese Academy of Sciences(IHEP-CAS) and are publicly available for download from the \textit{Insight}-HXMT website.
To process and fit the spectrum and obtain folded light curves, this research has made use of XRONOS and FTOOLS provided by NASA.

%\section*{References}

\appendix

\section{Different continuum models}
\begin{figure*}
	\includegraphics[scale=0.30]{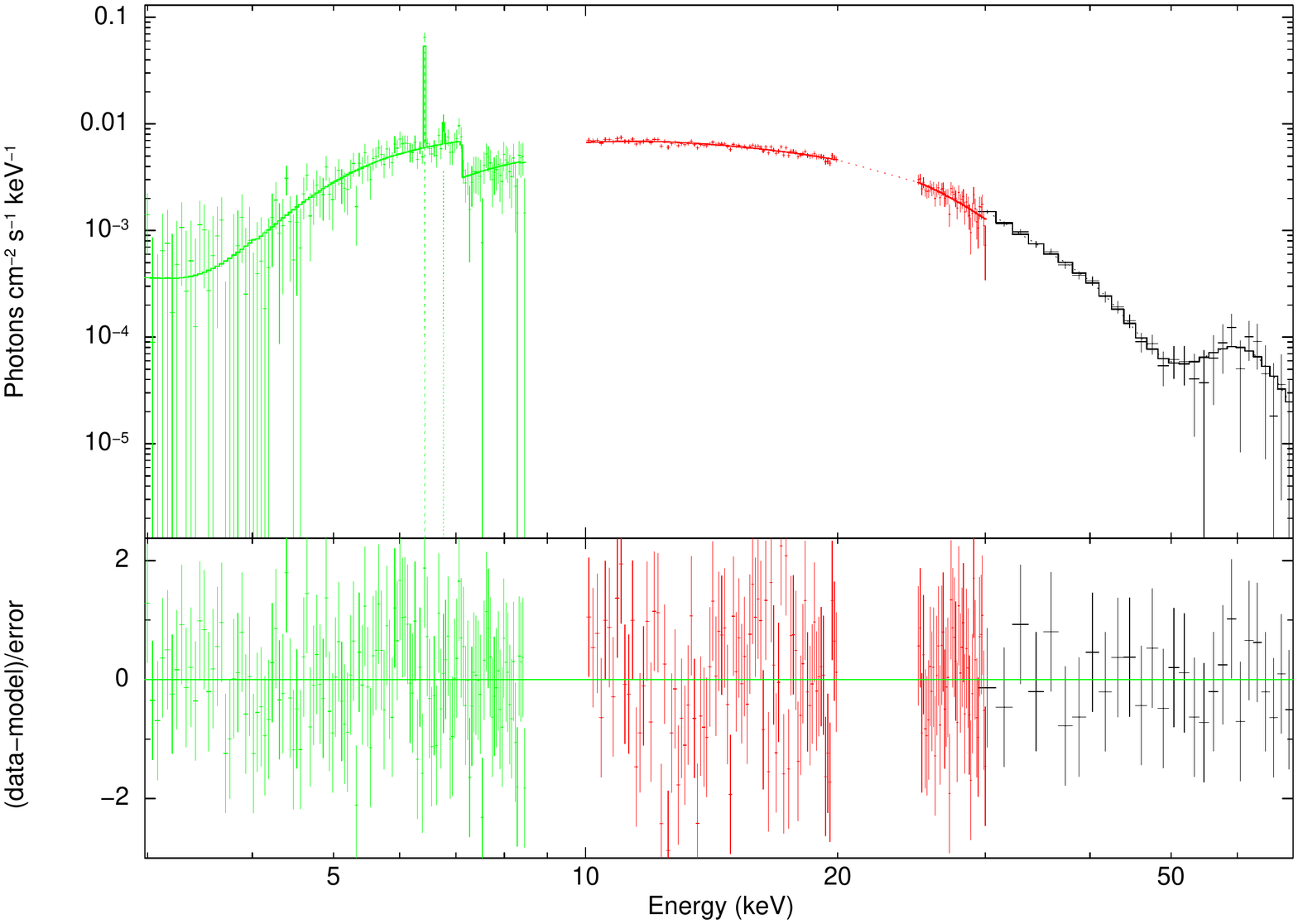}
	\includegraphics[scale=0.30]{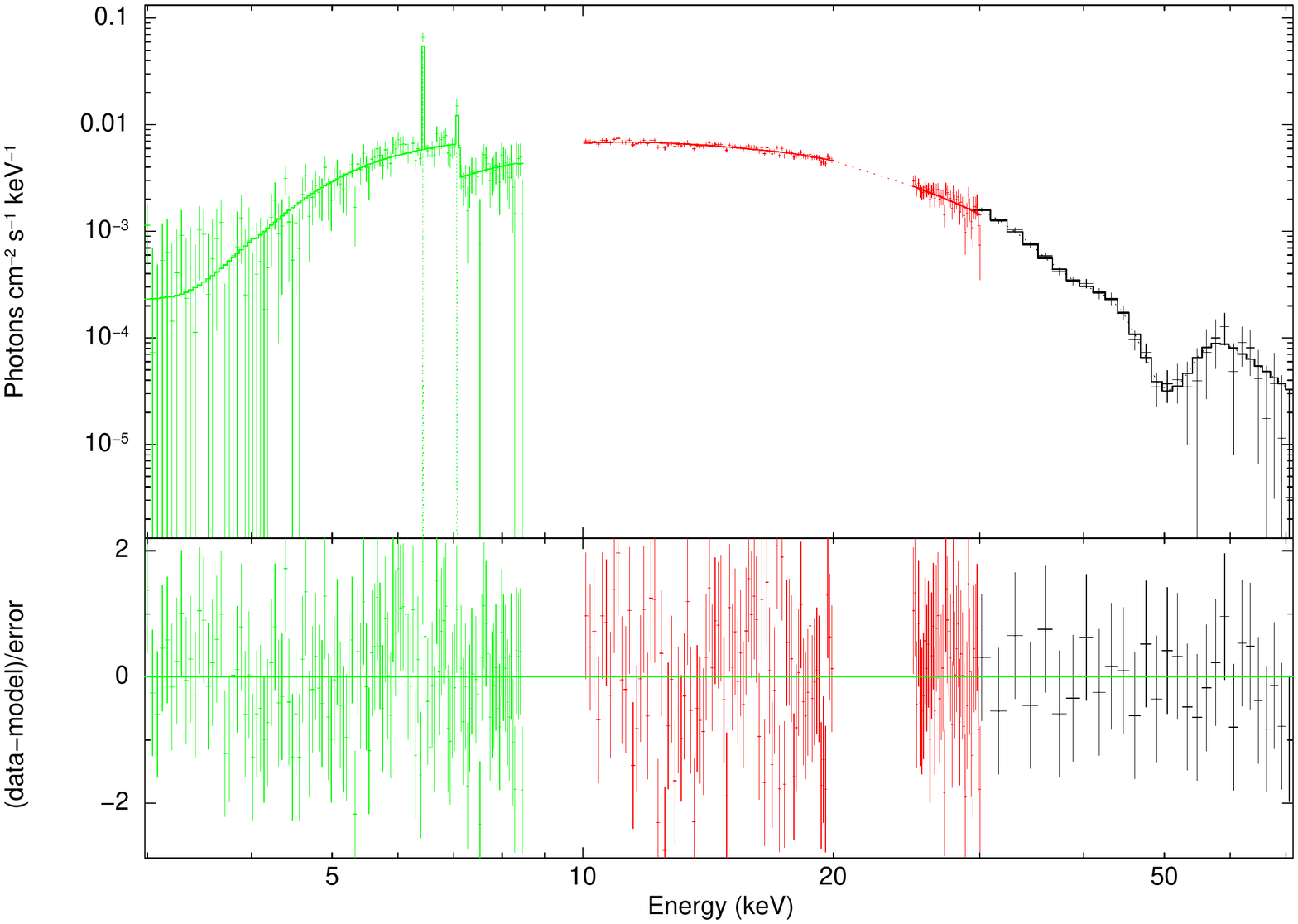}
	\includegraphics[scale=0.30]{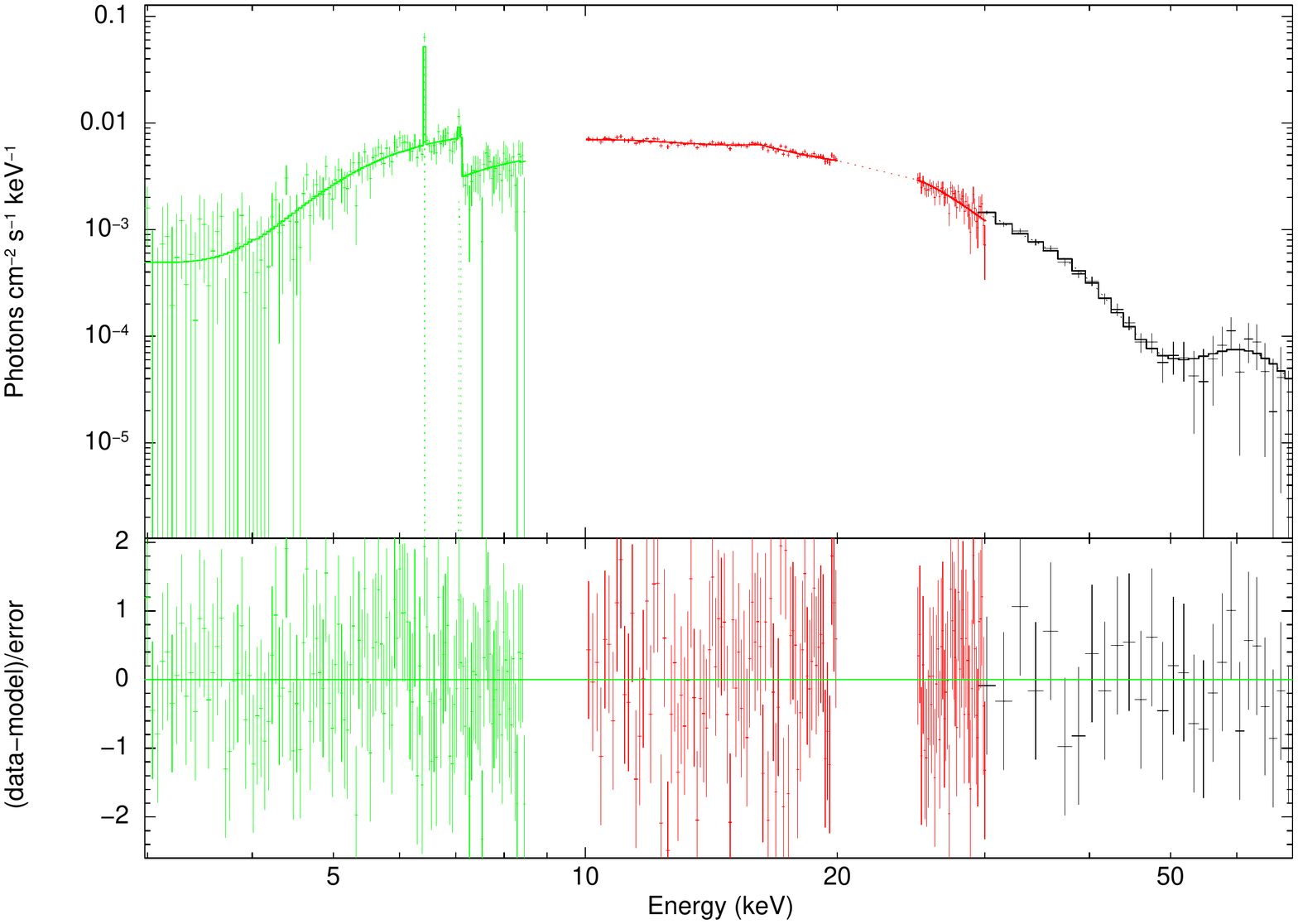}
	\includegraphics[scale=0.30]{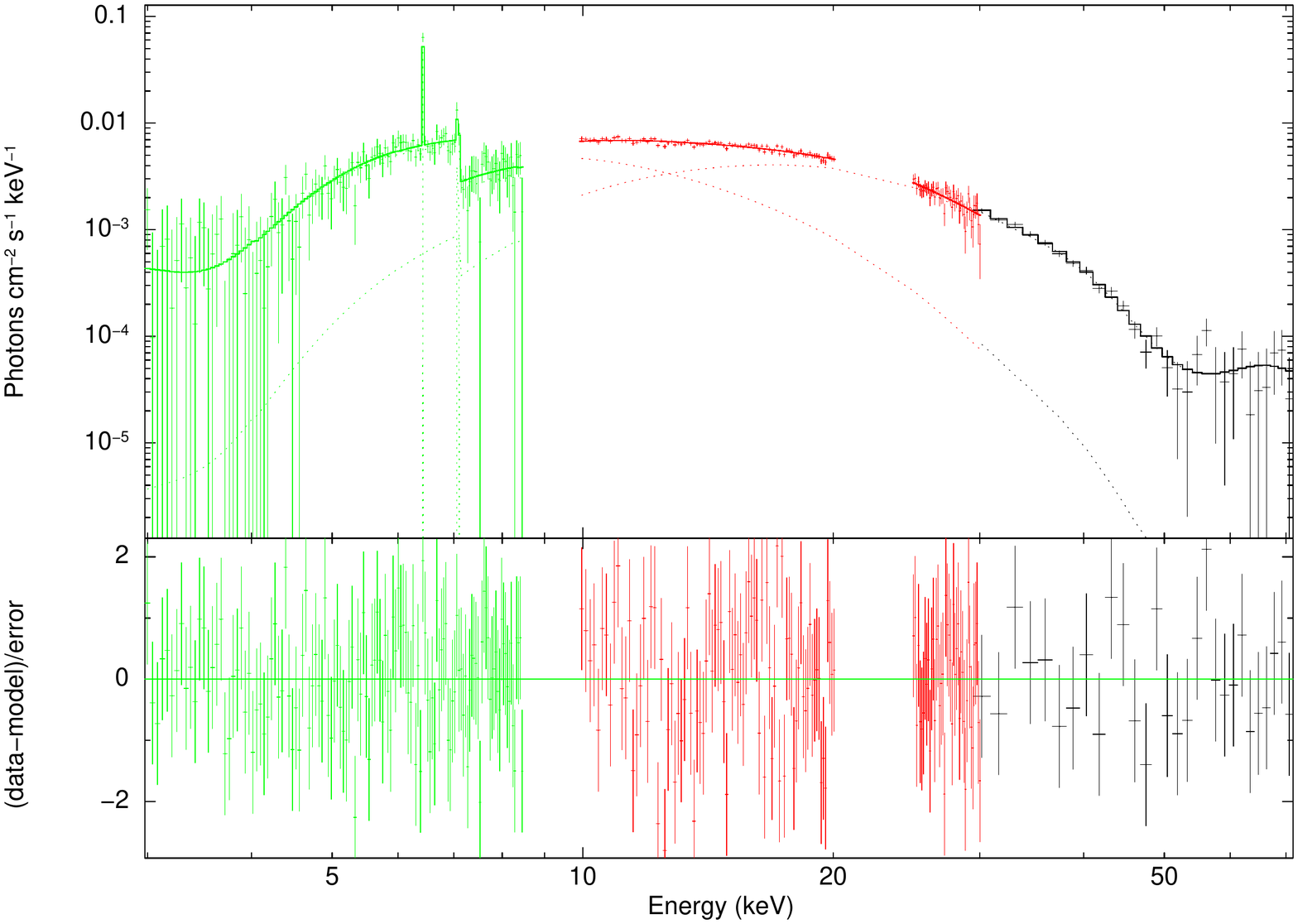}
	\label{fig:multispecmodel}
	\caption{Spectra of \gx in ObsID:P010130900701 from 3 --70 keV fitted with four different continuum models. Top left: FDcut; top right: newHcut; bottom left: highEcut; bottom right: NPEX.  Corresponding parameter values are shown in Table~\ref{tab:fourmodel}. All these continuum models with two Gabs can obtain an acceptable fit.}
\end{figure*}
\begin{table*}
	%\tabletypesize{\scriptsize}
	\scriptsize
	\caption{ Spectral parameters of \gx fitted with four different continuum models based on the \textit{Insight}-HXMT observations for the ObsID:P010130900701.}
	\label{tab:fourmodel}
	\renewcommand\arraystretch{1.8}
	\renewcommand\tabcolsep{3.0pt}
	% \setlength{\tabcolsep}{1.0mm}
	%\tablewidth{0pt}
	\begin{center}
		\begin{tabular}{l l l l l }
			\hline \hline
			Models & FDcut&newHcut&highEcut&NPEX\\
			\hline
			NH$_2$ ($10^{22}$ cm$^{-2}$) &	88.8$_{-5}^{+14}$&	$83_{-7}^{+10}$ &98$_{-19}^{+15}$ &101$_{-10}^{+18}$ \\
			PCF   &	$0.99_{-0.01}^{+0.01}$&$0.993_{-0.017}^{+0.006}$ &$0.983_{-0.017}^{+0.011}$&	$0.994_{-0.007}^{+0.006}$\\
			$\Gamma$  &	1.235$_{-0.017}^{+0.081}$& 1.21$_{-0.04}^{+0.10}$&	0.001$^{+0.158}_{-0.691}$ &1.30$_{-0.16}^{+0.13}$\\
			E$_{\rm fold}$ (keV) &	$4.87_{-0.88}^{+1.08}$&	$13.7_{-3.6}^{+1.4}$&	$6.6_{-1.3}^{+0.9}$	&-\\
			E$_{\rm cut}$ (keV) &	$49.9_{-7.5}^{+21.1}$&	$19.86_{-0.48}^{+3.01}$&	$16.16_{-0.46}^{+0.89}$	& 7.30$_{-0.41}^{+0.36}$\\
			E$_{\rm Fe K\alpha}$ (keV) & $6.435_{-0.026}^{+0.022}$& 6.435$_{-0.020}^{+0.017}$& $6.435_{-0.020}^{+0.013}$ &	$6.435_{-0.021}^{+0.017}$\\
			E$_{\rm Fe K\beta}$ (keV) &6.773$_{-5.58}^{+15.3}$ &7.1$_{-1.5}^{+1.0}$&7.08$_{-0.38}^{+0.13}$&7.08$_{-0.31}^{+0.66}$\\
			E$_{\rm cyc1}$ (keV) & 35.1$_{-1.8}^{+6.6}$	 &39.4$_{-3.1}^{+11}$&	31.7$_{-1.4}^{+2.7}$   &31.9$_{-1.4}^{+4.6}$ \\
			$\sigma_{\rm cyc1}$ (keV) & 6.2$_{-0.8}^{+2.8}$&4.1$_{-1.8}^{+7.6}$& $5.1_{-0.7}^{+0.9}$	&$5.7_{-0.6}^{+3.7}$\\
			Strength1 &	17.9$_{-8.3}^{+10.0}$ &6.5$_{-3.6}^{+14.8}$&	 $16.9_{-9.1}^{+18.5}$	& 7.6$_{-2.7}^{+7.4}$\\
			E$_{\rm cyc2}$ (keV) &	49.3$_{-1.7}^{+11.7}$&	$50.1_{-1.6}^{+29.6}$ &$46.9_{-1.8}^{+2.5}$	&	51.6$_{-1.6}^{+3.3}$ \\
			$\sigma_{\rm cyc2}$ (keV) &6.8$_{-0.5}^{+6.7}$&	3.2$_{-2.9}^{+10.9}$&	$8.6_{-2.0}^{+1.6}$ &	9.3$_{-2.8}^{+2.8}$\\
			Strength2 &	48.0$_{-7.5}^{+60.4}$ &	16$_{-15}^{+24}$ &	$59_{-16}^{+47}$ &	$49_{-16}^{+13}$\\
			Reduced-$\chi^2$ (dof) & 0.8646 (269) &	0.8645 (270)& 0.8323 (267)	 & 0.8732 (272)\\
			\hline \hline
		\end{tabular}
	\end{center}
\end{table*}

%%%%%%%%%%%%%%%%%%%%%%%%%%%%%%%%%%%%%%%%%%%%%%%%%%
% Don't change these lines
\bsp	% typesetting comment
\label{lastpage}

\begin{thebibliography}{}
\makeatletter
\relax
\def\mn@urlcharsother{\let\do\@makeother \do\$\do\&\do\#\do\^\do\_\do\%\do\~}
\def\mn@doi{\begingroup\mn@urlcharsother \@ifnextchar [ {\mn@doi@}
  {\mn@doi@[]}}
\def\mn@doi@[#1]#2{\def\@tempa{#1}\ifx\@tempa\@empty \href
  {http://dx.doi.org/#2} {doi:#2}\else \href {http://dx.doi.org/#2} {#1}\fi
  \endgroup}
\def\mn@eprint#1#2{\mn@eprint@#1:#2::\@nil}
\def\mn@eprint@arXiv#1{\href {http://arxiv.org/abs/#1} {{\tt arXiv:#1}}}
\def\mn@eprint@dblp#1{\href {http://dblp.uni-trier.de/rec/bibtex/#1.xml}
  {dblp:#1}}
\def\mn@eprint@#1:#2:#3:#4\@nil{\def\@tempa {#1}\def\@tempb {#2}\def\@tempc
  {#3}\ifx \@tempc \@empty \let \@tempc \@tempb \let \@tempb \@tempa \fi \ifx
  \@tempb \@empty \def\@tempb {arXiv}\fi \@ifundefined
  {mn@eprint@\@tempb}{\@tempb:\@tempc}{\expandafter \expandafter \csname
  mn@eprint@\@tempb\endcsname \expandafter{\@tempc}}}

\bibitem[\protect\citeauthoryear{Aasi et~al.,}{Aasi
  et~al.}{2015}]{aasi2015advanced}
Aasi J.,  et~al., 2015, Classical and quantum gravity, 32, 074001

\bibitem[\protect\citeauthoryear{{Abbott}, {Abbott}, {Abbott}  et~al.}{{Abbott}
  et~al.}{2017a}]{abbott17_gw170817_detection}
{Abbott} B.~P.,  {Abbott} R.,  {Abbott} T.~D.,   et~al., 2017a, \mn@doi [\prl]
  {10.1103/PhysRevLett.119.161101}, \href
  {http://adsabs.harvard.edu/abs/2017PhRvL.119p1101A} {119, 161101}

\bibitem[\protect\citeauthoryear{{Abbott}, {Abbott}, {Abbott}  et~al.}{{Abbott}
  et~al.}{2017b}]{abbott17_gw170817_Hubble}
{Abbott} B.~P.,  {Abbott} R.,  {Abbott} T.~D.,   et~al., 2017b, \mn@doi [\nat]
  {10.1038/nature24471}, \href
  {http://adsabs.harvard.edu/abs/2017Natur.551...85A} {551, 85}

\bibitem[\protect\citeauthoryear{{Abbott}, {Abbott}, {Abbott}  et~al.}{{Abbott}
  et~al.}{2017c}]{abbott17_gw170817_multimessenger}
{Abbott} B.~P.,  {Abbott} R.,  {Abbott} T.~D.,   et~al., 2017c, \mn@doi [\apj]
  {10.3847/2041-8213/aa91c9}, \href
  {http://adsabs.harvard.edu/abs/2017ApJ...848L..12A} {848, L12}

\bibitem[\protect\citeauthoryear{{Abbott}, {Abbott}, {Abbott}  et~al.}{{Abbott}
  et~al.}{2017d}]{abbott17_gw170817_gwgrb}
{Abbott} B.~P.,  {Abbott} R.,  {Abbott} T.~D.,   et~al., 2017d, \mn@doi [\apj]
  {10.3847/2041-8213/aa920c}, \href
  {http://adsabs.harvard.edu/abs/2017ApJ...848L..13A} {848, L13}

\bibitem[\protect\citeauthoryear{Abbott et~al.,}{Abbott
  et~al.}{2018a}]{abbott2018prospects}
Abbott B.~P.,  et~al., 2018a, Living Reviews in Relativity, 21, 3

\bibitem[\protect\citeauthoryear{{Abbott}, {Abbott}, {Abbott}  et~al.}{{Abbott}
  et~al.}{2018b}]{abbott_19_observing_scenarios}
{Abbott} B.~P.,  {Abbott} R.,  {Abbott} T.~D.,   et~al., 2018b, \mn@doi [Living
  Reviews in Relativity] {10.1007/s41114-018-0012-9}, \href
  {https://ui.adsabs.harvard.edu/abs/2018LRR....21....3A} {21, 3}

\bibitem[\protect\citeauthoryear{{Abbott}, {Abbott}, {Abbott}  et~al.}{{Abbott}
  et~al.}{2018c}]{abbott18_GW170817_NS_parameters}
{Abbott} B.~P.,  {Abbott} R.,  {Abbott} T.~D.,   et~al., 2018c, \mn@doi [\prl]
  {10.1103/PhysRevLett.121.161101}, \href
  {https://ui.adsabs.harvard.edu/abs/2018PhRvL.121p1101A} {121, 161101}

\bibitem[\protect\citeauthoryear{{Abbott}, {Abbott}, {Abbott}  et~al.}{{Abbott}
  et~al.}{2019}]{abbott19_O2_cosmo}
{Abbott} B.~P.,  {Abbott} R.,  {Abbott} T.~D.,   et~al., 2019, arXiv e-prints,
  \href {https://ui.adsabs.harvard.edu/abs/2019arXiv190806060T} {p.
  arXiv:1908.06060}

\bibitem[\protect\citeauthoryear{Abbott et~al.,}{Abbott
  et~al.}{2020a}]{abbott2020gw190425}
Abbott B.,  et~al., 2020a, arXiv preprint arXiv:2001.01761

\bibitem[\protect\citeauthoryear{{Abbott}, {Abbott}, {Abbott}  et~al.}{{Abbott}
  et~al.}{2020b}]{abbott19_EOS_model-select}
{Abbott} B.~P.,  {Abbott} R.,  {Abbott} T.~D.,   et~al., 2020b, \mn@doi
  [Classical and Quantum Gravity] {10.1088/1361-6382/ab5f7c}, \href
  {https://ui.adsabs.harvard.edu/abs/2020CQGra..37d5006A} {37, 045006}

\bibitem[\protect\citeauthoryear{{Abbott}, {Abbott}, {Abbott}  et~al.}{{Abbott}
  et~al.}{2020c}]{abbott20_GW190425}
{Abbott} B.~P.,  {Abbott} R.,  {Abbott} T.~D.,   et~al., 2020c, \mn@doi [\apjl]
  {10.3847/2041-8213/ab75f5}, \href
  {https://ui.adsabs.harvard.edu/abs/2020ApJ...892L...3A} {892, L3}

\bibitem[\protect\citeauthoryear{Abbott et~al.,}{Abbott
  et~al.}{2020d}]{abbott2020gw190814}
Abbott R.,  et~al., 2020d, The Astrophysical Journal Letters, 896, L44

\bibitem[\protect\citeauthoryear{Acernese et~al.,}{Acernese
  et~al.}{2014}]{acernese2014advanced}
Acernese F.,  et~al., 2014, Classical and Quantum Gravity, 32, 024001

\bibitem[\protect\citeauthoryear{{Aso} et~al.}{{Aso} et~al.}{2013}]{kagra}
{Aso} Y.,  et~al., 2013, \mn@doi [\prd] {10.1103/PhysRevD.88.043007}, \href
  {http://adsabs.harvard.edu/abs/2013PhRvD..88d3007A} {88, 043007}

\bibitem[\protect\citeauthoryear{{Biscoveanu}, {Vitale}  \&
  {Haster}}{{Biscoveanu} et~al.}{2019}]{biscoveanu2019b}
{Biscoveanu} S.,  {Vitale} S.,   {Haster} C.-J.,  2019, \mn@doi [\apjl]
  {10.3847/2041-8213/ab479e}, \href
  {https://ui.adsabs.harvard.edu/abs/2019ApJ...884L..32B} {884, L32}

\bibitem[\protect\citeauthoryear{{Biscoveanu}, {Thrane}  \&
  {Vitale}}{{Biscoveanu} et~al.}{2020}]{biscoveanu2019a}
{Biscoveanu} S.,  {Thrane} E.,   {Vitale} S.,  2020, \mn@doi [\apj]
  {10.3847/1538-4357/ab7eaf}, \href
  {https://ui.adsabs.harvard.edu/abs/2020ApJ...893...38B} {893, 38}

\bibitem[\protect\citeauthoryear{Buchner et~al.,}{Buchner
  et~al.}{2014}]{buchner2014x}
Buchner J.,  et~al., 2014, Astronomy \& Astrophysics, 564, A125

\bibitem[\protect\citeauthoryear{{Cantiello} et~al.,}{{Cantiello}
  et~al.}{2018}]{cantiello18}
{Cantiello} M.,  et~al., 2018, \mn@doi [\apjl] {10.3847/2041-8213/aaad64},
  \href {https://ui.adsabs.harvard.edu/abs/2018ApJ...854L..31C} {854, L31}

\bibitem[\protect\citeauthoryear{Capano et~al.,}{Capano
  et~al.}{2020}]{capano2020stringent}
Capano C.~D.,  et~al., 2020, Nature Astronomy, pp~1--8

\bibitem[\protect\citeauthoryear{Chopin \& Robert}{Chopin \&
  Robert}{2010}]{chopin2010properties}
Chopin N.,  Robert C.~P.,  2010, Biometrika, 97, 741

\bibitem[\protect\citeauthoryear{Collaboration, Collaboration
  et~al.}{Collaboration et~al.}{2020}]{ligo2020gw190412}
Collaboration L.~S.,  Collaboration V.,   et~al., 2020, arXiv preprint
  arXiv:2004.08342

\bibitem[\protect\citeauthoryear{Cornish \& Littenberg}{Cornish \&
  Littenberg}{2015}]{Cornish14}
Cornish N.~J.,  Littenberg T.~B.,  2015, \mn@doi [\cqg]
  {10.1088/0264-9381/32/13/135012}, 32, 135012

\bibitem[\protect\citeauthoryear{{Coughlin} \& {Dietrich}}{{Coughlin} \&
  {Dietrich}}{2019}]{coughlin19}
{Coughlin} M.~W.,  {Dietrich} T.,  2019, \mn@doi [\prd]
  {10.1103/PhysRevD.100.043011}, \href
  {https://ui.adsabs.harvard.edu/abs/2019PhRvD.100d3011C} {100, 043011}

\bibitem[\protect\citeauthoryear{Dalcin, Paz, Kler  \& Cosimo}{Dalcin
  et~al.}{2011}]{dalcin2011parallel}
Dalcin L.~D.,  Paz R.~R.,  Kler P.~A.,   Cosimo A.,  2011, Advances in Water
  Resources, 34, 1124

\bibitem[\protect\citeauthoryear{{Farah} et~al.,}{{Farah}
  et~al.}{2019}]{farah19}
{Farah} W.,  et~al., 2019, \mn@doi [\mnras] {10.1093/mnras/stz1748}, \href
  {https://ui.adsabs.harvard.edu/abs/2019MNRAS.488.2989F} {488, 2989}

\bibitem[\protect\citeauthoryear{Farr et~al.,}{Farr
  et~al.}{2016}]{farr2016parameter}
Farr B.,  et~al., 2016, The Astrophysical Journal, 825, 116

\bibitem[\protect\citeauthoryear{Goldstein et~al.,}{Goldstein
  et~al.}{2017}]{goldstein2017ordinary}
Goldstein A.,  et~al., 2017, The Astrophysical Journal Letters, 848, L14

\bibitem[\protect\citeauthoryear{{Graham} et~al.,}{{Graham}
  et~al.}{2020}]{graham2020}
{Graham} M.~J.,  et~al., 2020, \mn@doi [\prl] {10.1103/PhysRevLett.124.251102},
  \href {https://ui.adsabs.harvard.edu/abs/2020PhRvL.124y1102G} {124, 251102}

\bibitem[\protect\citeauthoryear{{Hernandez Vivanco}, {Smith}, {Thrane}  \&
  {Lasky}}{{Hernandez Vivanco} et~al.}{2019a}]{hernandezvivanco2019b}
{Hernandez Vivanco} F.,  {Smith} R.,  {Thrane} E.,   {Lasky} P.~D.,  2019a,
  \mn@doi [\prd] {10.1103/PhysRevD.100.043023}, \href
  {https://ui.adsabs.harvard.edu/abs/2019PhRvD.100d3023H} {100, 043023}

\bibitem[\protect\citeauthoryear{{Hernandez Vivanco}, {Smith}, {Thrane},
  {Lasky}, {Talbot}  \& {Raymond}}{{Hernandez Vivanco}
  et~al.}{2019b}]{hernandezvivanco2019}
{Hernandez Vivanco} F.,  {Smith} R.,  {Thrane} E.,  {Lasky} P.~D.,  {Talbot}
  C.,   {Raymond} V.,  2019b, \mn@doi [\prd] {10.1103/PhysRevD.100.103009},
  \href {https://ui.adsabs.harvard.edu/abs/2019PhRvD.100j3009H} {100, 103009}

\bibitem[\protect\citeauthoryear{Metzger}{Metzger}{2017}]{metzger2017kilonovae}
Metzger B.~D.,  2017, Living reviews in relativity, 20, 3

\bibitem[\protect\citeauthoryear{{Mooley} et~al.,}{{Mooley}
  et~al.}{2018}]{mooley18}
{Mooley} K.~P.,  et~al., 2018, \mn@doi [\nat] {10.1038/s41586-018-0486-3},
  \href {https://ui.adsabs.harvard.edu/abs/2018Natur.561..355M} {561, 355}

\bibitem[\protect\citeauthoryear{{Morisaki} \& {Raymond}}{{Morisaki} \&
  {Raymond}}{2020}]{morisaki2020}
{Morisaki} S.,  {Raymond} V.,  2020, arXiv e-prints, \href
  {https://ui.adsabs.harvard.edu/abs/2020arXiv200709108M} {p. arXiv:2007.09108}

\bibitem[\protect\citeauthoryear{{\"O}zel \& Freire}{{\"O}zel \&
  Freire}{2016}]{ozel2016masses}
{\"O}zel F.,  Freire P.,  2016, Annual Review of Astronomy and Astrophysics,
  54, 401

\bibitem[\protect\citeauthoryear{Perego, Radice  \& Bernuzzi}{Perego
  et~al.}{2017}]{perego20172017gfo}
Perego A.,  Radice D.,   Bernuzzi S.,  2017, The Astrophysical Journal Letters,
  850, L37

\bibitem[\protect\citeauthoryear{{Singer} \& {Price}}{{Singer} \&
  {Price}}{2016}]{singer16a}
{Singer} L.~P.,  {Price} L.~R.,  2016, \mn@doi [\prd]
  {10.1103/PhysRevD.93.024013}, 93, 024013

\bibitem[\protect\citeauthoryear{{Singer} et~al.}{{Singer}
  et~al.}{2016}]{singer16b}
{Singer} L.~P.,  et~al., 2016, \mn@doi [\apj] {10.3847/2041-8205/829/1/L15},
  829, L15

\bibitem[\protect\citeauthoryear{Skilling}{Skilling}{2004}]{skilling2004nested}
Skilling J.,  2004, in AIP Conference Proceedings. pp 395--405

\bibitem[\protect\citeauthoryear{Skilling}{Skilling}{2006}]{Skilling06}
Skilling J.,  2006, \mn@doi [Bayesian Analysis] {10.1214/06-BA127}, 1, 833

\bibitem[\protect\citeauthoryear{Smith \& Ashton}{Smith \&
  Ashton}{2019}]{smith2019expediting}
Smith R.,  Ashton G.,  2019, arXiv preprint arXiv:1909.11873

\bibitem[\protect\citeauthoryear{Smith, Field, Blackburn, Haster, P{\"{u}}rrer,
  Raymond  \& Schmidt}{Smith et~al.}{2016}]{smith16}
Smith R.,  Field S.~E.,  Blackburn K.,  Haster C.-J.,  P{\"{u}}rrer M.,
  Raymond V.,   Schmidt P.,  2016, \mn@doi [\prd] {10.1103/PhysRevD.94.044031},
  94

\bibitem[\protect\citeauthoryear{{Speagle}}{{Speagle}}{2020}]{dynesty}
{Speagle} J.~S.,  2020, \mn@doi [\mnras] {10.1093/mnras/staa278}, \href
  {https://ui.adsabs.harvard.edu/abs/2020MNRAS.493.3132S} {493, 3132}

\bibitem[\protect\citeauthoryear{Valenti et~al.,}{Valenti
  et~al.}{2017}]{valenti2017discovery}
Valenti S.,  et~al., 2017, The Astrophysical Journal Letters, 848, L24

\bibitem[\protect\citeauthoryear{{Veitch} et~al.}{{Veitch}
  et~al.}{2015}]{veitch15}
{Veitch} J.,  et~al., 2015, \mn@doi [\prd] {10.1103/PhysRevD.91.042003}, 91,
  042003

\bibitem[\protect\citeauthoryear{Yang et~al.,}{Yang
  et~al.}{2017}]{yang2017empirical}
Yang S.,  et~al., 2017, The Astrophysical Journal Letters, 851, L48

\bibitem[\protect\citeauthoryear{You, Zhu, Ashton, Thrane  \& Zhu}{You
  et~al.}{2020}]{you2020standard}
You Z.-Q.,  Zhu X.-J.,  Ashton G.,  Thrane E.,   Zhu Z.-H.,  2020, arXiv
  preprint arXiv:2004.00036

\bibitem[\protect\citeauthoryear{{Zackay}, {Dai}  \& {Venumadhav}}{{Zackay}
  et~al.}{2018}]{zackay18}
{Zackay} B.,  {Dai} L.,   {Venumadhav} T.,  2018, arXiv e-prints, \href
  {https://ui.adsabs.harvard.edu/abs/2018arXiv180608792Z} {p. arXiv:1806.08792}

\makeatother
\end{thebibliography}


\begin{thebibliography}{300}	
	
\bibitem[{Abarr, et al.}{2020}]{2020ApJ...891...70A} Abarr Q., et al., 2020, ApJ, 891, 70	
\bibitem[{Burderi}{2000}]{Burderi2000} {Burderi} L., {Di Salvo} T., {Robba} N.~R., {La Barbera} A.,  {Guainazzi} M., 2000, ApJ, 530, 429

\bibitem[{Becker, et al.}{2012}]{Becker2012} {Becker P. A.} et al., 2012, A\&A 544, A123

\bibitem[Bailer-Jones, et al.]{2018} Bailer-Jones C. A. L., Rybizki J., Fouesneau, M., Mantelet, G., Andrae R.,  2018, AJ, 156, 58
%\bibitem[{Borkus, et al.}{1998}]{Borkus1998} Borkus V.~V., et al., 1998, AstL, 24, 60	
	
%\bibitem[{Caballero{et~al.}(2008)}]{Caballero2008} Caballero, I., Santangelo, A.,
%		 Kretschmar, P., {et~al.} 2008, A\&A, 480, L17
\bibitem[{cao et al.}{2020}]{cao2020} Cao X. et al., 2020, Science China Physics, Mechanics, and Astronomy, 63, 249504
\bibitem[{chen et al.}{2020}]{chen2020}Chen Y., Cui, W.,  Li W. et al. 2020, Science China Physics, Mechanics, and Astronomy, 63, 249505
	
\bibitem[{Doroshenko, et al.}{2010}]{Doroshenko2010} Doroshenko V., Santangelo A., Suleimanov V., Kreykenbohm I., Staubert R., Ferrigno C., Klochkov D., 2010, A\&A, 515, A10

\bibitem[{Endo, et al.}{2002}]{Endo2002} Endo T., Ishida M., Masai K., Kunieda H., Inoue H., Nagase F., 2002, ApJ, 574, 879

%\bibitem[{Evangelista, et al.}{2010}]{Evangelista2010} Evangelista Y., et al., 2010, ApJ, 708, 1663

\bibitem[{F{\"u}rst, et al.}{2011}]{Fuerst2011} F{\"u}rst F. et al., 2011, A\&A, 535, A9

\bibitem[{F{\"u}rst, et al.}{2018}]{Fuerst2018} F{\"u}rst F. et al., 2018, A\&A, 620, A153
\bibitem[{Gimenez-Garcia, et al.}{2015}]{Garcia2015} Gimenez-Garcia A. et al., 2015, A\&A, 576, A108
%\bibitem[{G{\"o}{\u{g}}{\"u}{s}, Kreykenbohm \& Belloni}{2011}]{Gogus2011}
%        G{\"o}{\u{g}}{\"u}s E., Kreykenbohm I., Belloni T.~M., 2011, A\&A, 525, L6
\bibitem[{harding}{1984}]{harding1984}Harding A. K., Kirk J. G., Galloway D. J., M\'esz\'aros P. et al. 1984, ApJ, 278, 369
\bibitem[{Haberl}{1991}]{Haberl1991} Haberl F., 1991, ApJ, 376, 245
\bibitem[{Ikhsanov \& Finger}{2012}]{Ikhsanov2012} Ikhsanov N.~R., Finger M.~H., 2012, ApJ, 753, 1
%\bibitem[{{\.I}nam \& Baykal}{2000}]{Inam2000} {\.I}nam S. {\c{C}}., Baykal A., 2000, A\&A, 353, 617
\bibitem[{Islam \& Paul}{2014}]{Islam2014} Islam N., Paul B., 2014, MNRAS, 441, 2539

\bibitem[Ji, L. et al.]{Ji2021} Ji L. et al., 2021, MNRAS, 501, 2522 
%\bibitem[{Islam \& Paul}{2018}]{Islam2018} Islam N., Paul B., 2018, cosp, 42, E1.13-40-18, cosp...42

%\bibitem[{Islam}{2019}]{Islam2019} Islam N., 2019, IAUS, 346, 59, IAUS 346

%\bibitem[{Kaper, et al.}{1995}]{Kaper1995} Kaper L., Lamers H.~J.~G.~L.~M.,
%         Ruymaekers E., van den Heuvel E.~P.~J., Zuiderwijk E.~J., 1995, A\&A, 300, 446

\bibitem[{Kaper, van der Meer \& Najarro}{2006}]{Kaper2006} Kaper L., van der Meer A., Najarro F., 2006, A\&A, 457, 595

%\bibitem[{Kawai, et al.}{1985}]{Kawai1985} Kawai N., Makishima K., Matsuoka M., Mitani K.,
%         Murakami T., Nagase F., 1985, PASJ, 37, 647

%\bibitem[{Kelley, Rappaport \& Petre}{1980}]{Kelley1980} Kelley R., Rappaport S.,
%         Petre R., 1980, ApJ, 238, 699

\bibitem[{Koh, et al.}{1997}]{Koh1997} Koh D.~T., et al., 1997, ApJ, 479, 933

\bibitem[{Kreykenbohm, et al.}{2004}]{Kreykenbohm2004} Kreykenbohm I., et al., 2004, A\&A, 427, 975

\bibitem[{La Barbera, et al.}{2005}]{Barbera2005} La Barbera A., Segreto A., Santangelo A.,
        Kreykenbohm I., Orlandini M., 2005, A\&A, 438, 617	

%\bibitem[{Leahy, et al.}{1988}]{Leahy1988} Leahy D.~A., Nakajo M., Matsuoka M., Kawai N.,
%         Koyama K., Makino F., 1988, PASJ, 40, 197

%\bibitem[{Leahy, et al.}{1989a}]{Leahy1989a} Leahy D.~A., Matsuoka M., Kawai N., Makino F., 1989a, MNRAS, 236, 603

%\bibitem[{Leahy, et al.}{1989b}]{Leahy1989b} Leahy D.~A., Matsuoka M., Kawai N., Makino F., 1989b, MNRAS, 237, 269

%\bibitem[{Leahy \& Matsuoka}{1990}]{Leahy1990} Leahy D.~A., Matsuoka M., 1990, ApJ, 355, 627
\bibitem[{Larsson}{1996}]{Larsson1996} Larsson S., 1996, A\&A, 117, 197

\bibitem[{Leahy}{1987}]{Leahy1987} Leahy D. A., 1987, A\&A, 180, 275

\bibitem[{Leahy}{1991}]{Leahy1991} Leahy D.~A., 1991, MNRAS, 250, 310

\bibitem[{Leahy}{2002}]{Leahy2002} Leahy D.~A., 2002, A\&A, 391, 219

\bibitem[{Leahy \& Kostka}{2008}]{Leahy2008} Leahy D.~A., Kostka M., 2008, MNRAS, 384, 747

%\bibitem[{Leahy, et al.}{1985}]{Leahy1985} Leahy D.~A., Kawai N., Koyama K., Makino F.,
%         Matsuoka M., 1985, JRASC, 79, 240

%\bibitem[{Lewin, et al.}{1971}]{Lewin1971} Lewin W.~H.~G., McClintock J.~E.,
%         Ryckman S.~G., Smith W.~B., 1971, ApJL, 166, L69
\bibitem[{li et al. }{2020}]{li2020} Li X., Tan Y. et al. 2020, Journal of High Energy Astrophysics, 27, 64
\bibitem[{Liu, et al.}{2018}]{Liu2018} Liu J., Soria R., Qiao E., Liu J., 2018, MNRAS, 480, 4746
\bibitem[{liu et al. }{2020}]{liu2020} Liu C. et al. 2020, Science China Physics, Mechanics, and Astronomy, 63, 249503

%\bibitem[{Maurer, et al.}{1980}]{Maurer1980} Maurer G.~S., Johnson W.~N., Kurfess J.~D.,
%         Strickman M.~S., 1980, BAAS, 12, 513
\bibitem[{M\'esz\'aros}{1992}]{Mészáros1992} M\'esz\'aros, P. 1992, High-energy Radiation from Magnetized Neutron Stars (Chicago: University of Chicago Press)

\bibitem[{mihara1995}]{mihara1995} Mihara, T. 1995, PhD thesis, Univ. of Tokyo

\bibitem[{Makishima \& Mihara}{ 1992}]{maki1992} Makishima, K., \& Mihara, T. 1992, Magnetic Fields of Neutron Starss



\bibitem[{M{\"o}nkk{\"o}nen, et al.}{2020}]{2020MNRAS.494.2178M} M{\"o}nkk{\"o}nen J., Doroshenko V., Tsygankov S.~S., Nabizadeh A., Abolmasov P., Poutanen J., 2020, MNRAS, 494, 2178

\bibitem[{Mukherjee \& Paul}{2004}]{Mukherjee2004} Mukherjee U., Paul B., 2004, A\&A, 427, 567
\bibitem[mushtukov2015]{mushtukov2015a} Mushtukov A., Suleimanov V.~F., Tsygankov S.~S., Poutanen J., 2015, MNRAS, 447, 1847–1856
\bibitem[{Nabizadeh, et al.}{2019}]{Nabizadeh2019} Nabizadeh A., M{\"o}nkk{\"o}nen J.,
        Tsygankov S.~S., Doroshenko V., Molkov S.~V., Poutanen J., 2019, A\&A, 629, A101
        
\bibitem[{Nishimura}{2015}]{Nishimura2015} Nishimura O., 2015, ApJ, 807, 164
%\bibitem[{Orlandini \& Morfill}{1992}]{Orlandini1992} Orlandini M., Morfill G.~E., 1992, ApJ, 386, 703
\bibitem[{Orlandini, et al.}{1998}]{Orlandini1998} {Orlandini} M. et al., 1998, ApJ, 500, 163-166
         
\bibitem[{Orlandini, et al.}{2000}]{Orlandini2000} Orlandini M., dal Fiume D., Frontera F., Oosterbroek T., Parmar A.~N., Santangelo A., Segreto A., 2000, AdSpR, 25, 417

\bibitem[{Pravdo et~al.}{1995}]{Pravdo1995} Pravdo S.~H, Day C.~S.~R.,
	     Angelini L. et al., 1995, ApJ, 454, 872

\bibitem[{Pravdo \& Ghosh}{2001}]{Pravdo2001} Pravdo S.~H., Ghosh P., 2001, ApJ, 554, 383

%\bibitem[Ricker et al.(1973)]{Ricker1973} Ricker, G.~R., McClintock,
%         J.~E., Gerassimenko, M., et al.\ 1973, ApJ, 184, 237

%\bibitem[{Rothschild, Soong \& Worrall}{1983}]{Rothschild1983} Rothschild R.~E., Soong Y.,
%        Worrall D.~M., 1983, BAAS, 15, 940

%\bibitem[{Rothschild \& Soong}{1987}]{Rothschild1987} Rothschild R.~E., Soong Y., 1987, ApJ, 315, 154

\bibitem[{Sato, et al.}{1986}]{Sato1986} Sato N., Nagase F., Kawai N., Kelley R.~L., Rappaport S., White N.~E., 1986, ApJ, 304, 241

%\bibitem[{Saraswat, et al.}{1996}]{Saraswat1996} Saraswat P., et al., 1996, ApJ, 463, 726
	
%\bibitem[{Servillat, et al.}{2014}]{Servillat2014A} Servillat M., Coleiro A., Chaty S.,
%         Rahoui F., Zurita Heras J.~A., 2014, ApJ, 797, 114

\bibitem[{Staubert et al.}{2019}]{Staubert2019} Staubert R. et al. 2019, A\&A, 622, A61
\bibitem[{Suchy, et al.}{2012}]{Suchy2012} Suchy S., et al., 2012, ApJ, 745, 124

\bibitem[{Schwarm, et al.}{2017}]{Schwarm2017} Schwarm F.-W, et al., 2017, A\&A, 601, A99

%\bibitem[{Swank, et al.}{1976}]{Swank1976} Swank J.~H., Becker R.~H., Boldt E.~A., Holt S.~S.,
%         Pravdo S.~H., Rothschild R.~E., Serlemitsos P.~J., 1976, ApJL, 209, L57

\bibitem[{Tanaka}{1987}] Tanaka Y., 1986, IAU Colloq. 89: Radiation Hydrodynamics in Stars , Compact Objects, 255, 198

\bibitem[{Torrejon et al.}{2010}]{Torrejon2010} Torrejon J. M., Schulz N. S., Nowak M. A., \& Kallman T. R., 2010, ApJ, 715,
947

\bibitem[{Tsygankov, et al.}{2004}]{Tsygankov2004} Tsygankov S.~S., Lutovinov A.~A., Grebenev S.~A., Gilfanov M.~R., Sunyaev R.~A., 2004, AstL, 30, 540
\bibitem[{Wang}{2013}]{wang13} Wang, W., 2013, MNRAS, 432, 954
\bibitem[{Wang}{2014}]{wang14} Wang, W., 2014, Research in Astronomy and Astrophysics, 14, 565
%\bibitem[{Watanabe, et al.}{2003}]{Watanabe2003} Watanabe S., et al., 2003, ApJL, 597, L37

\bibitem[{White \& Swank}{1984}]{White1984} White N.~E., Swank J.~H., 1984, ApJ, 287, 856
\bibitem[{Yu \& Wang}{2016}]{yuwei16} Yu G. W., Wang W., 2016, Astronomical Research \& Technology, 13, 11
\bibitem[{zhang et al.}{2020}]{zhang2020} Zhang S.-N. et al. 2020, Science China Physics, Mechanics, and Astronomy, 63, 249502
\bibitem[{Zheng, Liu \& Gou}{2020}]{Zheng2020} Zheng X., Liu J., Gou L., 2020, MNRAS, 491, 4802
\end{thebibliography}
\end{document}